\documentclass{ieeeaccess}
\usepackage{cite}
\usepackage{amsmath,amssymb,amsfonts}
\usepackage{algorithmic}
\usepackage{graphicx,epsfig}
\usepackage{caption}
\usepackage{subcaption}
\usepackage{textcomp}
\usepackage{amsmath,amssymb,amsfonts}
\usepackage{graphicx}
\usepackage{textcomp,epsfig,epstopdf}
\usepackage{caption,setspace}
\def\BibTeX{{\rm B\kern-.05em{\sc i\kern-.025em b}\kern-.08em
    T\kern-.1667em\lower.7ex\hbox{E}\kern-.125emX}}
\begin{document}
\history{Date of publication xxxx 00, 2020, date of current version February 24, 2020.}
\doi{10.1109/ACCESS.2019.DOI}

\title{BER Performance of Spatial Modulation Systems Under a Non-Stationary Massive MIMO Channel Model}
\author{
\uppercase{Yu Fu}\authorrefmark{1}, 
\uppercase{Cheng-Xiang Wang}\authorrefmark{1,2,3} \IEEEmembership{Fellow,~IEEE}, \uppercase{Xuming Fang}\authorrefmark{4} \IEEEmembership{Senior Member,~IEEE}, \uppercase{Li Yan}\authorrefmark{4},  \uppercase{Stephen McLaughlin}\authorrefmark{1} \IEEEmembership{Fellow,~IEEE}}
\address[1]{Institute of Sensors, Signals and Systems, School of Engineering and Physical Sciences, Heriot-Watt University, Edinburgh, EH14 4AS, U.K. (e-mail: \{y.fu, cheng-xiang.wang, s.mclaughlin\}@hw.ac.uk)}
\address[2]{National Mobile Communications Research Laboratory, School of Information Science and Engineering, Southeast University, Nanjing, 210096, China (e-mail: chxwang@seu.edu.cn)}
\address[3]{Purple Mountain Laboratories, Nanjing, 211111, China}
\address[4]{The Key Lab of Info Coding and Transmission, Southwest Jiaotong University, Chengdu 610031, China (e-mail: xmfang@swjtu.edu.cn, liyan12047001@my.swjtu.edu.cn)}
%\tfootnote{This paragraph of the first footnote will contain support information, including sponsor and financial support acknowledgment. For example, ``This work was supported in part by the U.S. Department of Commerce under Grant BS123456.''}

\markboth
{Y. Fu \headeretal: Preparation of Papers for IEEE Access}
{Y. Fu \headeretal: Preparation of Papers for IEEE Access}

\corresp{Corresponding author: Cheng-Xiang Wang (e-mail: cheng-xiang.wang@hw.ac.uk).}
\tfootnote{This work was supported by the National Key R\&D Program of China under Grant 2018YFB1801101, the National Natural Science Foundation of China (NSFC) under Grant 61960206006, the EPSRC TOUCAN project under Grant EP/L020009/1, the High Level Innovation and Entrepreneurial Talent Introduction Program in Jiangsu, the Research Fund of National Mobile Communications Research Laboratory, Southeast University, under Grant 2020B01, the Fundamental Research Funds for the Central Universities under Grant 2242019R30001, the Huawei Cooperation Project, and the EU H2020 RISE TESTBED2 project under Grant 872172. This work was partly presented in Dr. Yu Fu's Ph.D. thesis, Sept. 2015.}

\begin{abstract}
In this paper, the bit error rate (BER) performance of spatial modulation (SM) systems is investigated both theoretically and by simulation in a non-stationary Kronecker-based massive multiple-input-multiple-output (MIMO) channel model in multi-user (MU) scenarios. Massive MIMO SM systems are considered in this paper using both a time-division multiple access (TDMA) scheme and a block diagonalization (BD) based precoding scheme, for different system settings. Their performance is compared with a vertical Bell labs layered space-time (V-BLAST) architecture based system and a conventional channel inversion system. It is observed that a higher cluster evolution factor can result in better BER performance of SM systems due to the low correlation among sub-channels. Compared with the BD-SM system, the SM system using the TDMA scheme obtains a better BER performance but with a much lower total system data rate. The BD-MU-SM system achieves the best trade-off between the data rate and the BER performance among all of the systems considered. When compared with the V-BLAST system and the channel inversion system, SM approaches offer advantages in performance for MU massive MIMO systems. 
\end{abstract}


\titlepgskip=-15pt

\maketitle

\section{Introduction}
Fifth generation (5G) wireless communication systems are accelerating towards their final deployment. In order to satisfy the dramatically increasing demands of reliability, capacity, energy efficiency and spectrum efficiency required for 5G wireless communication systems many novel physical layer transmission technologies have been widely investigated \cite{CXW14Comag5G}. In recent years, massive multiple-input-multiple-output (MIMO) technology which employs tens to hundreds of antennas has attracted significant research interest due to its potential capability to fulfill the aforementioned  requirements\cite{Larsson14ComagMassive, Rusek14SigmagMassive, Mietzner09ComSurvey, Lu14JSAC}, and consequently has been treated as one of the key technologies of 5G wireless systems \cite{5GJSAC}.

In practice, how to efficiently implement a massive MIMO system is still a challenging problem. Existing MIMO technologies, including spatial multiplexing and spatial diversity \cite{Gebert03JsacSTBC, FoschiniBellVblast}, have achieved significant successes in current fourth generation (4G) wireless communication systems \cite{Rusek14SigmagMassive}. However, these technologies suffer from several drawbacks such as  inter-channel interference (ICI) and high system complexity, which make these technologies unsuitable for massive MIMO systems. Spatial modulation (SM) has attracted considerable research interest due to offering a trade-off between system performance and complexity \cite{Renzo11MagSM, SM1, Renzo14ProcSMSurvey}.  This advantage suggests SM as a potential solution for implementation of massive MIMO systems.

The basic concept of SM was first introduced in \cite{SM1}. Essentially at each time instant, only one antenna is active. The index of the active antenna is a resource available to carry information. As a result of this extra resource, SM can obtain a relatively high spectral efficiency with low-order modulation schemes\cite{Renzo11MagSM, Renzo14ProcSMSurvey}. As only one antenna is used to transmit at each time instant, the ICI problem is avoided and simple receiver (Rx) algorithms can be employed, significantly reducing the system complexity \cite{SM2}. In \cite{Review1,YuThesis}, the performance of SM was studied in different wireless communication scenarios such as the vehicle-to-vehicle scenario, high-speed train scenario, and the massive MIMO scenario. As an innovative method to convey information, the principle of SM has been further developed and applied to different resource domains. For example, generalized SM was proposed in \cite{GSM2,PiyaGSM2}, employing a subgroup of antennas to transmit signals and can further improve the system efficiency compared with conventional SM. In \cite{Ind1, PiyaIndex1}, index modulation (IM) was introduced, which applied the concept of SM to the index of subcarriers in orthogonal frequency division multiplexing (OFDM) systems. In \cite{PraSM}, a practical implementation of SM using a hardware testbed was demonstrated. The concept of SM is deployed as the generalized beamspace modulation in \cite{mmWaveSM1,mmWaveSM2}, for mmWave massive MIMO systems. 
In addition to conventional radio frequency wireless communications SM has also been used in optical wireless communication systems \cite{optSM1,optSM2} where light emitting diode (LED) devices rather than antennas are used to transmit data. Good surveys on the range of SM methods and applications can be found in \cite{SMsurvey1,SMsurvey2}.

In realistic conditions, a massive MIMO system always works in the multi-user (MU) regime. The question of how to eliminate inter-user-interference (IUI) is an important technical question. The first approach is to use a time-division multiple access (TDMA) scheme \cite{ZhouEurasip11TDMU, Ngo13AACTD}, which means all antennas serve only one user in a particular time slot. In the TDMA scheme, the IUI problem is avoided, as channel knowledge of the transmitter (Tx) is not required and the system complexity is low. However, as it uses only one active antenna to transmit data to one user, when the TDMA scheme is combined with a SM system the utilization of antenna elements and the system achievable sum rate are lower than competing approaches. Alternatively, to enable the Tx to send signals to all users at the same time using all transmit antennas precoding methods are employed . In \cite{SpencerCommag04Precod, JohamTransSig05Precod, PeelTVT05Precod}, research for precoding schemes for MU-MIMO systems were reported, in which some well-known precoding schemes such as minimum mean-squared error (MMSE), zero-forcing (ZF), and matched filter (MF) methods have been investigated. However, none of these methods are suitable for SM systems as they all eliminate the channel's impact, losing the data carried by the antenna indexes \cite{NarayananWCNC14MUSM}. In addition, all of these methods only support the use of a single receive antenna for each user, i.e. ignoring the benefit offered from  receive diversity. In \cite{Spencer04BD}, a method named block diagonalization (BD) was introduced, the basic idea was to compose the precoding matrix that represents the null-space of IUI by applying singular value decomposition (SVD). The most attractive characteristic of this method is that it can
support multiple receive antennas with the ability to retain the channel's impact, making it possible to establish a precoding SM system with receive diversity gain.   

In a massive MIMO system, as a large number of antennas are distributed over a wide spatial range, different antenna elements can observe different sets of clusters \cite{Wang185GSur, Wang16China}. Measurement results in \cite{Payami12ECAP, Gao12} have demonstrated this observation. Geometry-based stochastic models proposed in \cite{WuJSAC14Massive, WuTWC15Massive,WuITC18MA} are the first series of channel models that consider the evolution of clusters, accurately capturing the characteristics of massive MIMO channels. However, as a result of the high computational complexity, it is not easy to use these channel models for further theoretical analysis. In this paper, a Kronecker-based stochastic model (KBSM) for massive MIMO scenarios proposed in \cite{WuICCC15KBSM} is used to overcome this difficulty. This channel model considers the evolution of clusters for each antenna element using a survival property matrix. This matrix is an abstraction of the birth-death process of clusters. As this channel model is a KBSM, it is easier to use for the required theoretical analysis.

Most existing research studies to date, on SM technologies, only focused on single-user point-to-point scenarios \cite{VVTVT16, Fuvtc14SMVV, RenzoTVT12SMBER, FuTWC16SMHST, YounisVTC13SM, ZhangTWC2014SM} for small-scale MIMO systems. In recent years, some researchers have attempted to investigate SM in MU massive MIMO scenarios, related work has been reported in \cite{RenzoTVT11BERmuSSK , KadirTVT13OFDMsm, HumadiHadawi14SMMU, YangTSP12SM,WangICCT12SMpre}. However, there are still some research gaps in the research of MU-SM systems. First, most research on MU-SM only focuses on the uplink scenario, only limited reported research such as \cite{Review2} exists for the downlink case. Second, some researchers have tried to use precoding schemes to eliminate the IUI, but in most existing work except for the BD precoding technique, each user can use only a single receive antenna because of the limitation of selected precoding methods. The benefit of receive diversity is omitted. Third, there has not been sufficient research on practical massive MIMO systems. In some of the research reported on massive MIMO systems the number of transmit antennas was less than 32. In addition, for most reported research, only conventional MIMO channel models or identically independent distributed (i.i.d.) channel models are used. These channel models do not accurately represent the fading characteristics of massive MIMO channels.

In order to begin addressing these gaps, in this work the performance of SM is investigated for a KBSM which was specifically developed for MU massive MIMO scenarios \cite{YuThesis}. In this paper, massive MIMO SM systems are established for both a TDMA scheme and for a BD based precoding scheme. Different channel configurations and system settings are considered to comprehensively evaluate the system performance of both approaches. 

The main contributions of this paper are as follows.
\begin{enumerate}
\item SM is investigated in a range of MU scenarios using a KBSM that was specifically proposed for massive MIMO systems. A comprehensive study on the BER performance of SM systems in this massive MIMO channel model is carried out. How different channel factors, such as the evolution factors of clusters and the number of users, affect the system BER performance are thoroughly studied. 

\item  An accurate theoretical BER expression for MU-SM system using a BD-precoding scheme is presented. To the best of authors' knowledge, there is no published study to derive the theoretical BER expression of MU-SM systems with a precoding scheme supporting multiple receive antennas for each user. 

\item The BER performance of different MIMO technologies including TDMA-SM, BD-SM, BD-vertical Bell labs layered space-time architecture (V-BLAST), and conventional channel inversion \cite{Spencer04MUMIMO} are compared. The influence of different system settings on the system performance such as the number of users and data rate are investigated. This is the first time that SM using a BD precoding scheme is compared with other MIMO technologies in MU massive MIMO scenarios.
 
\end{enumerate}
The remainder of the paper is organized as follows. In Section II, the system architectures of the MU-SM systems in a massive MIMO scenario are introduced. In Section III, the KBSM for massive MIMO systems is described. The theoretical expression for the BER for SM in a MU massive scenarios is derived in Section IV. In Section V, simulation results are presented and analyzed. Finally, conclusions are drawn in Section VI.        

\section{Systems architecture}
\subsection{The TDMA-MU-SM system}

\begin{figure}[t]
\includegraphics[width=3.4in]{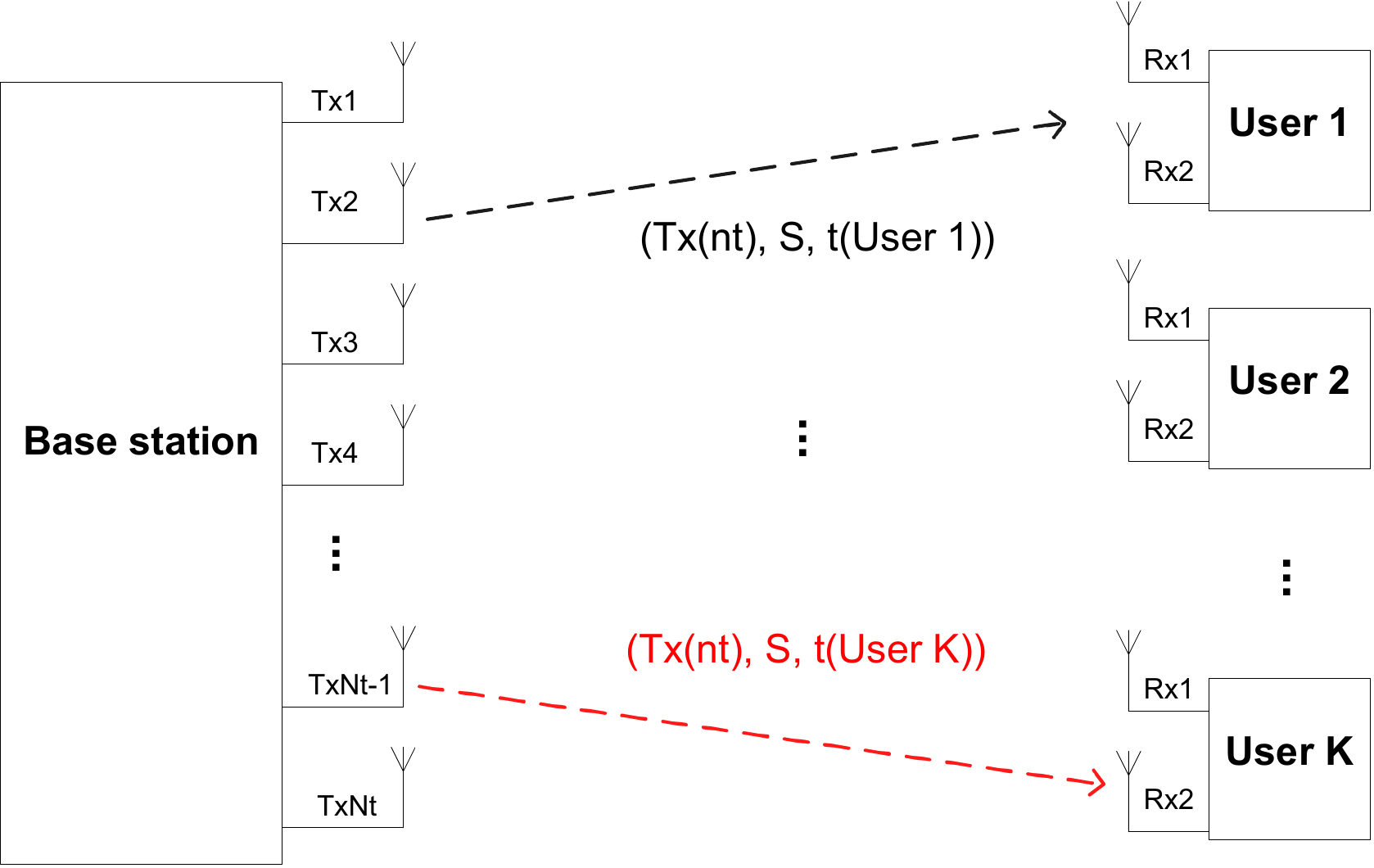}
\caption{Block diagram of a TDMA-MU-SM system.}
\label{TDblock}
%\label{fig_sim}
\end{figure}

In a MU system using the TDMA scheme, at each time instant, the base station (BS) only transmits data to one user. Thus, compared with other multiple access schemes, the system complexity of the TDMA system can be significantly decreased as it is not necessary to employ complex precoding methods to mitigate IUI. Let us assume there is a system with $K$ users, the BS is equipped with $N_t$ antenna elements. At each instant, only one antenna is used to transmit data to a single user. The bit stream for each user is divided into two parts, the first ${\log _2}({N_t})$ bits are carried by the antenna index, which is also used for the antenna selection. The other ${\log _2}({M})$ bits are transmitted through the digital modulated symbol, where $M$ is the modulation order. Thus at each time instant, ${m\ =\ {\log _2}({N_t})+{\log _2}(M)}$ bits are transmitted, where ${m}$ is the spectral efficiency of the SM system in the unit of bits/symbol. An example of a TDMA-MU-SM system is shown in Fig. \ref{TDblock}, where at the time instant for a selected user, the modulated symbol $S$ is transmitted through the selected active antenna to the user.  

Mathematically, the detailed expression for the received signal in $K$ users TDMA system at the time slot of  the $k$-th user can be written as  
\begin{equation}\label{rxvec}
\boldsymbol{Y}(t_k) = \boldsymbol{H} \boldsymbol{S}(t_k) + \boldsymbol{N}, k=1.2,...,K
\end{equation}
where $\boldsymbol{H}$ is the channel matrix vector and can be expressed as
\begin{equation}\label{Huser} 
{\boldsymbol{H}} = {\left[ {\begin{array}{*{20}{c}}
  \boldsymbol{H_{1}^T}&\boldsymbol{H_{2}^T}&{...}&\boldsymbol{H_{k }^T\begin{array}{*{20}{c}}
  {...}&\boldsymbol{H_{K}^T} 
\end{array}} 
\end{array}} \right]^T}.
\end{equation}
Each element $\boldsymbol{H_k} $($1 \leqslant k \leqslant K$) is the channel matrix of the $k$-th user, $\boldsymbol{N}$ is the complex additive white Gaussian noise (AWGN) vector for the system, $[.]^{\rm{T}}$ denotes the transpose operation, $\boldsymbol{S}(t_k)$ is the transmit vector for $K$ users at the time slot for $k$-th user $t_k$, while each element is the signal vector for the corresponding user. As the system is operating a TDMA scheme, at the time instant of the $k$-th user, only the element of the $k$-th user in $\boldsymbol{S}(t_k)$ is not zero and can be expressed as
\begin{equation}\label{Kuser} 
\boldsymbol{S} = {[\begin{array}{*{20}{c}}
  0&{...}&{\boldsymbol{S_k}}&{...}&0 
\end{array}]_{K \times 1}^{\rm{T}}}.
\end{equation}
 
Let $\boldsymbol{S_k}(t_k)$ denote the transmit signal vector of the $k$-th user at the Tx side at its allocated time $t_k$, which is a $N_t \times 1$-size vector. For a TDMA-SM system, as only one transmit antenna is active at each time instant, $\boldsymbol{S_k}(t_k)$ can be further expressed as $\boldsymbol{S_k}(t_k) = {\left[ {0,...,s,...,0} \right]_{ {N_t}}}^T$.
As each user is equipped with $N_r$ antennas, at the Rx side of the $k$-th user, the received $N_r \times 1$-dimensional signal vector $ {y_k}(t_k)$ can be written as
\begin{equation}\label{rxvec2}
\boldsymbol{y_k}(t_k) = \boldsymbol{H_k}(t_k) \boldsymbol{S_k}(t_k) + \boldsymbol{n}.
\end{equation}
In addition, $\boldsymbol{H_k}(t_k)$ is the ${N_r} \times {N_t}$ channel matrix of the $k$-th user at time $t_k$, each element ${{h_{q,p}}(t_k)}$ is the channel impulse response between the $p$-th Tx antenna and the $q$-th antenna Rx, $\boldsymbol{n}$ is the AWGN vector. The detailed procedure of generating channel impulse responses for the massive channel model will be introduced in the next section.

In this system, the Tx and the Rx are assumed to be in perfect synchronization in both time and frequency domains. It is also assumed that full channel state information (CSI) is available at the Rx in this paper. The optimum maximum likelihood (ML) detector proposed in \cite{SM2} is employed to estimate the transmit antenna index $\hat p$ and the transmitted data symbol ${ s}$ for each user as
\begin{equation}\label{Sm1}
\begin{array}{c
}
\left[ {\hat p,{ s}} \right]_k = \arg \mathop {\min }\limits_{p,m} \left( {\left\| {\boldsymbol{y_k} - \boldsymbol{H_k{S_{l,m}}}} \right\|_F^2} \right)\\
p \in \left\{ {1:{N_T}} \right\}{\kern 1pt} {\kern 1pt} \text{and} {\kern 1pt} {\kern 1pt} {\kern 1pt} m \in \{ 1:M\}
\end{array}
\end{equation}
where ${\left\| \cdot \right\|_F}$ denotes the Frobenius norm.      

\subsection{The BD-MU-SM system}
\begin{figure}
 \hspace{-0.2cm} \includegraphics[width=3.4 in]{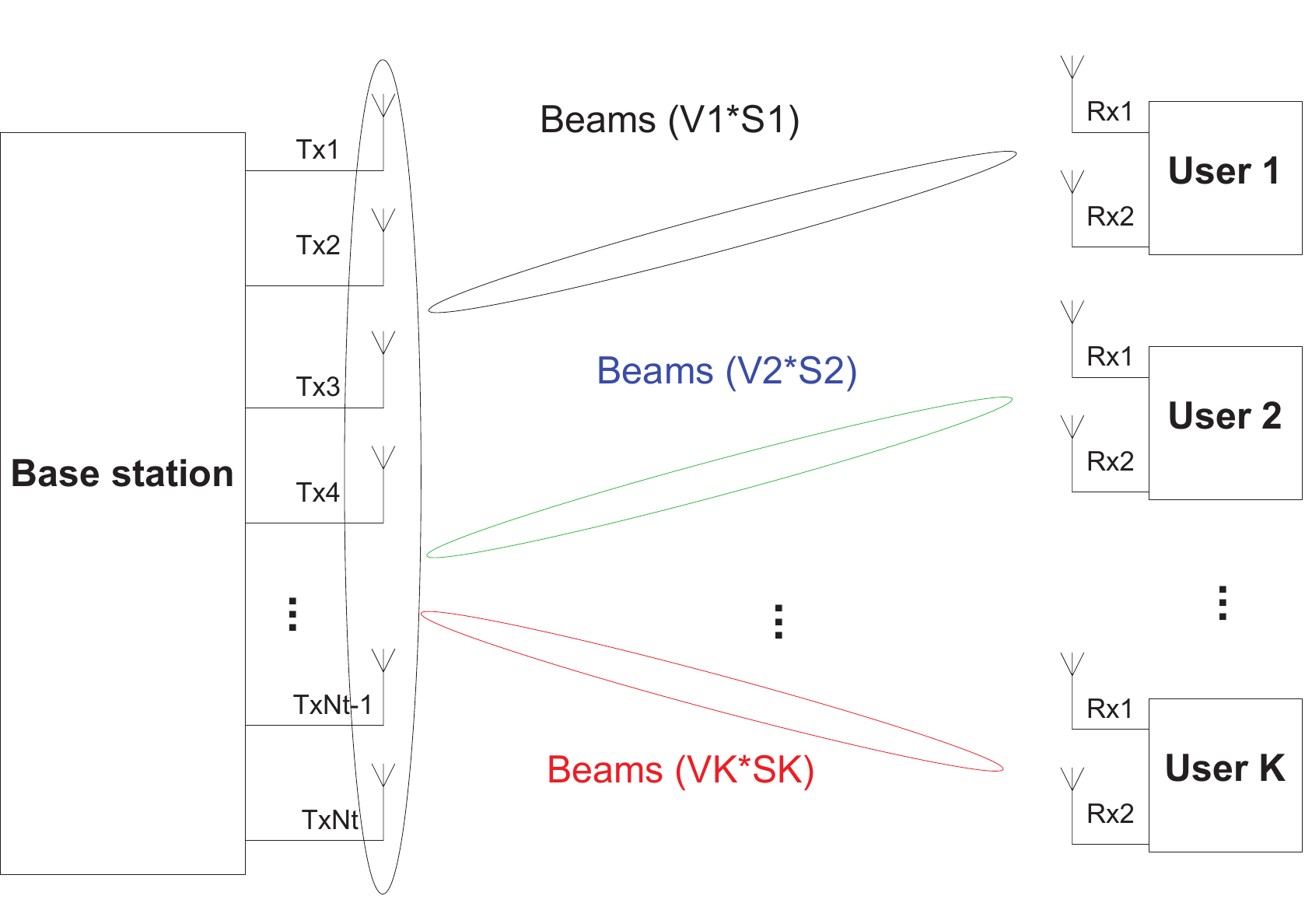}
\caption{Block diagram of a BD-MU-SM system.}
\label{BDblock}
%\label{fig_sim}
\end{figure}

In Fig. \ref{BDblock}, the system architecture of a BD precoding MU-SM system is shown. 
The principle of BD precoding was introduced in \cite{Spencer04BD}, whose basic idea is to generate the precoding matrix from the null space of the IUI. In this work and \cite{YuThesis}, we build the BD-MU-SM system with this method.

Let us still assume a system has $K$ users where each user is equipped with $N_r$ receive antennas, while the BS uses $N_t$ antennas to serve all the $K$ users at the same time. Thus, the channel matrix can still be expressed by (\ref{Huser}). For the $k$-th user ($1 \leqslant k \leqslant K$), its interference channel matrix $\tilde H$ can be defined as
\begin{equation}\label{Hint} 
\boldsymbol{{\tilde H}_k} = {\left[ {\begin{array}{*{20}{c}}
  \boldsymbol{H_1^T}&{...}&\boldsymbol{H_{k - 1}^T}&\boldsymbol{H_{k + 1}^T\begin{array}{*{20}{c}}
  {...}&\boldsymbol{H_{K}^T} 
\end{array}} 
\end{array}} \right]^T}.
\end{equation}
The precoding matrix for this user can be obtained through the singular decomposition (SVD) of  ${\tilde H}_k$  as:
\begin{equation}\label{HSVD} 
\boldsymbol{\tilde H_k} = \boldsymbol{\tilde U_k}{\tilde \Sigma _k}{\left[ {\begin{array}{*{20}{c}}
  \boldsymbol{\tilde V_k^{(1)}}&\boldsymbol{\tilde V_k^{(0)}} 
\end{array}} \right]^H}
\end{equation}
where ${[.]^H}$ is the Hermitian transpose and $\boldsymbol{\tilde V_k^{(0)}}$ is the precoding matrix we need. When we use the BD based precoding scheme, it is easy to prove that $\boldsymbol{H_j}\boldsymbol{\tilde V_k^{(0)}} = 0,\forall j \ne k$. Thus, the IUI is eliminated.

In order to clearly describe the principle, let us define $J_k=N_t-(K-1)\times N_r$, which is the maximum number of beamforming  patterns of the system and it should be a positive number. Thus, for a BD based MU system, it should firstly fulfill the condition that  ${N_t} > \sum\limits_{k = 1}^{K - 1} {Nr}$. The precoding matrix ${\tilde V_k^{(0)}}$ is with the dimension of $N_t \times J_k$. In a conventional SM system, the antenna index is used to carry information. However, for a BD-MU-SM system, it uses the BD precoding scheme to generate beams to eliminate IUI and all antennas are active. Thus, in a BD-SM system, what is used to perform SM is the beamforming pattern index rather than the antenna index. For a BD-SM system, the following conditions should be fulfilled. First, in order to use the beam pattern index to carry information, the minimum pattern numbers should be not less than 2 as $J_k \geqslant 2$. Secondly, the maximum number of bits that can be carried by the pattern index is $\left\lfloor {\rm{log{_2}}({J_k})} \right\rfloor$,  where $\left\lfloor {} \right\rfloor$ is the floor function. We can observe that in a BD-MU-SM system, the pattern number of each user is $2\leqslant N\leqslant J_k $, the precoding matrix ${\tilde V}_k$ has the dimension  $N_t \times N$. The spectral efficiency $m$ can be expressed as ${m\ =\ {\log_2}({N}) + {\log_2}(M)}$.
Let us define the precoded transmitted signal vector $\boldsymbol{S}$ as
\begin{equation}\label{CodedSig} 
\boldsymbol{S}=\left[ {\begin{array}{*{20}{c}}{{{\tilde V}_1}\boldsymbol{S_1}}&{...}&{{{\tilde V}_k}\boldsymbol{S_k}}&{...}&{{{\tilde V}_K}\boldsymbol{S_K}}\end{array}} \right]
\end{equation}
where $\boldsymbol{S_k}(1 \leqslant k \leqslant K)$ is the transmitted signal vector for the $k$-th user with the dimension of  $N \times1$ and there is only one non-zero element in  the vector $\boldsymbol{S_k}$.
For the system, this procedure can mathematically be expressed as
\begin{align}\label{CodedMat} 
 \boldsymbol{Y}& \hspace{-0.06cm}=\hspace{-0.06cm} \left[\hspace{-0.06cm} {\begin{array}{*{20}{c}}
  {{y_1}} \\ 
   \vdots  \\ 
  {{y_k}} \\ 
   \vdots  \\ 
  {{y_K}} 
\end{array}}\hspace{-0.06cm} \right]\hspace{-0.06cm}= \hspace{-0.06cm}\left[ \hspace{-0.06cm}{\begin{array}{*{20}{c}}
  {{H_1}} \\ 
   \vdots  \\ 
  {{H_k}} \\ 
   \vdots  \\ 
  {{H_K}} 
\end{array}} \hspace{-0.06cm}\right] \hspace{-0.03cm}{\left[\hspace{-0.06cm} {\begin{array}{*{20}{c}}
  {{{\tilde V}_1}\boldsymbol{S_1}}& \hspace{-0.1cm}\cdots\hspace{-0.1cm} &{{{\tilde V}_k}\boldsymbol{S_k}}& \hspace{-0.1cm}\cdots\hspace{-0.1cm} &{{{\tilde V}_K}\boldsymbol{S_K}} 
\end{array}} \hspace{-0.06cm}\right]}\hspace{-0.06cm}+\hspace{-0.06cm} \boldsymbol{n} \nonumber\\ &=\left[ {\begin{array}{*{20}{c}}
  {{H_1}{{\tilde V}_1}\boldsymbol{S_1}}& \cdots &0& \cdots &0 \\ 
   \vdots & \ddots & \cdots & \cdots & \vdots  \\ 
  0& \cdots &{{H_k}{{\tilde V}_k}\boldsymbol{S_k}}& \cdots &0 \\ 
   \vdots & \cdots & \cdots & \ddots & \vdots  \\ 
  0& \cdots &0& \cdots &{{H_K}{{\tilde V}_K}\boldsymbol{S_K}} 
\end{array}} \right] +  \boldsymbol{n}.
\end{align}

It can be observed from (\ref{CodedMat}) that IUI can be avoided. For each user, its $N_r \times 1$ received signal vector can be expressed as
\begin{equation}\label{RecVec} 
\boldsymbol{y_k} = {H_k}{\rho _k}{{\tilde V}_k}\boldsymbol{S_k} + n
\end{equation}
where ${\rho _k}$ is the scaling factor to normalize the power and can be expressed as 
\begin{equation}\label{Norm} 
{\rho _k} = \sqrt {\frac{E_{T_r}}{tr({({{\tilde V}_k}) \cdot {{({{\tilde V}_k})}^H}})}}
\end{equation}
where $E_{T_r}$ is the transmit power and $tr(.)$ is the trace of a matrix.

In BD based precoding schemes, the effective channel can be expressed as ${H_E} = H\tilde V$. In this channel, there are $N$ beamforming patterns generated which also equals to the dimension of the precoding matrix. If we investigate the cross-correlation function (CCF) of different beamforming patterns $p_i$ and $p_k$, we can write
\begin{equation}\label{CCF} 
CCF_{{p_i}{p_k}} = (H\tilde V(i)){(H\tilde V(k))^*} = H\tilde V(i)\tilde V{(k)^*}{H^*}.
\end{equation}

From the SVD process, it is easy to prove that ${\tilde V(i)} $ and $\tilde V(k)$ are orthogonal. Thus, the CCF of two beam patterns can be derived to be zero, which means there is no correlation between two beamforming patterns. In this paper, beamforming patterns can be treated as sub-channels in conventional SM systems. It is well-known that a low correlation between sub-channels can greatly improve the performance of SM systems. Thus, the advantage of using the BD based precoding scheme is that it can offer i.i.d. propagation channels for spatial modulation systems even in highly correlated channel conditions (for example, the Tx and Rx are close to each other, which is the usual scenario for millimeter wave communications systems).

The ML detector still works in the BD framework. However, the task of the detector is to estimate the beamforming pattern number $p$ rather than the antenna index, the transmitted symbol $s$ is jointly detected at the same time, which can be expressed as
\begin{equation}\label{ML_BD}
\begin{array}{c
}
\left[ {\hat p,{ s}} \right]_k = \arg \mathop {\min }\limits_{p,l} \left( {\left\| {\boldsymbol{y_k} - {H_k}{{{\tilde V}_k}}\boldsymbol{S_{p, m}}} \right\|_F^2} \right)\\
p \in \left\{ {1:N} \right\}{\kern 1pt} {\kern 1pt} \text{and} {\kern 1pt} {\kern 1pt} {\kern 1pt} m \in \{ 1:M\}
\end{array}.
\end{equation} 

For this utilized ML detector, its detection complexity can be calculated following the method mentioned in\cite{Review2}. Please note, in our system, as some information is carried by the beam index rather the active antenna index. By considering (\ref{ML_BD}), the complexity of the ML dector can be expressed as
\begin{align}
\delta_{BDSM(ML)}=&2{N_t}{N_r}{J_k}+2N({N_r}^2)2^{{N_r}{\log_2(M)}}\\\nonumber
&+2NM{N_R}+NM.   
\end{align}

\section{The massive MIMO channel model}\label{S2}
For a conventional KBSM MIMO channel model, the general expression is, 
\begin{equation}\label{KBSM1} 
H = R_R^{_2^1}{H_W}R_T^{_2^1}
\end{equation}
where $H_W$ is usually assumed as the i.i.d. Gaussian entries with the dimension of $N_r \times N_t$ , $R_R$ ($N_r \times N_r$ dimension) and $R_T$ ($N_t \times N_t$ dimension) are the correlation matrices of Tx and Rx receptively, and can be assumed as Toeplitz matrices when using uniform linear antennas. The Tx and Rx correlation matrices are composed by spatial correlation coefficients ${T_{R/T,mn}}$, which indicate the correlations between the $m$-th and the $n$-th antenna. 
As mentioned in the previous sections, due to the wide spread of antennas in massive MIMO systems, different cluster sets may be observed by different antennas. The evolution of the clusters on the array axis is expressed as the scatter survival probability with the expression of \cite{WuICCC15KBSM}
\begin{equation}\label{KBSM2} 
{E_{R/T,mn}} = {e^{ - \beta \left| {m - n} \right|}}
\end{equation}
which describes the probability that the $m$-th antenna shares the same clusters with the $n$-th antenna, $\beta$ the evolution factor which indicates how fast the scatters disappear. At the Rx, a similar analysis can also be carried out.  The correlation coefficient can be expressed as
\begin{equation}\label{KBSM3} 
R{_{R/T,mn}^E} = {E_{R/T,mn}}{R_{R/T,mn}}.
\end{equation}
Finally, the channel impulse response can be generated based on (\ref{KBSM1}) and (\ref{KBSM3}) as in \cite{WuICCC15KBSM}
\begin{equation}\label{KBSMf} 
H = {(R_R^E)^{\frac{1}{2}}}{H_W}{(R_T^E)^{\frac{1}{2}}} = {({E_R} \odot {R_R})^{\frac{1}{2}}}{H_W}{({E_T} \odot {R_T})^{\frac{1}{2}}}
\end{equation}
where $\odot$ is the Hadamard product of two matrices.
\begin{figure}[t]
 \hspace{-0.2cm} \includegraphics[width=3.5 in]{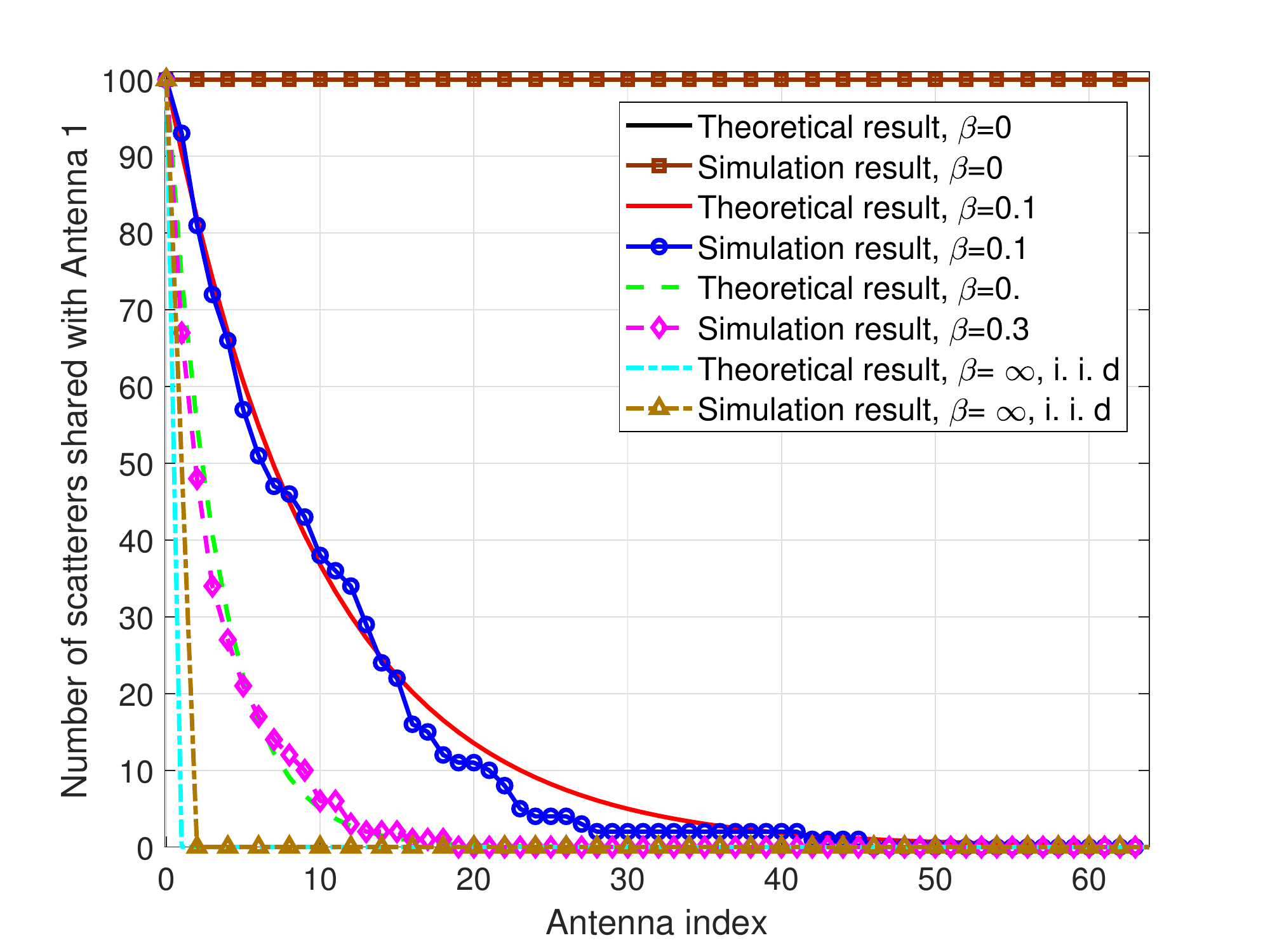}
\caption{Number of scatters shared with Antenna 1 in terms of antenna indices (100 initial scatters observed in Antenna 1) \cite{WuICCC15KBSM}.} 
\label{ClusterShare}
\end{figure}
In Fig. \ref{ClusterShare}, the number of scatters that other antennas shared with antenna 1 is shown considering different cluster evolution factors. It is necessary to note that when $\beta=0$, all antennas share the same set of clusters, thus the channel model becomes a conventional MIMO channel model. When $\beta=\infty$, the channel becomes an i.i.d. channel model as no cluster is shared by different antennas. 

\begin{figure}[t]
\includegraphics[width=3.5 in]{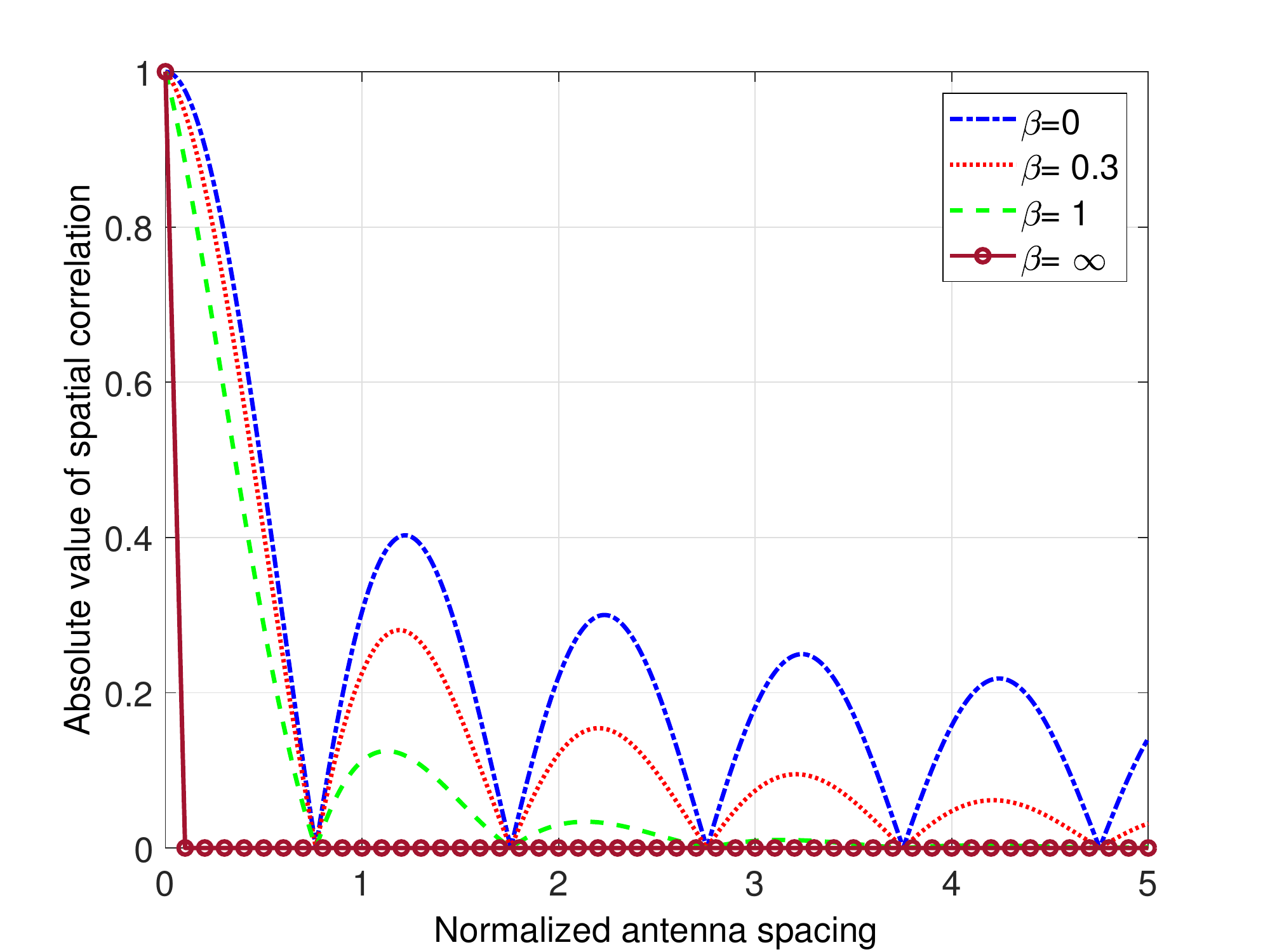}
\caption{The spatial correlation of the KBSM for massive MIMO systems considering different evolution factor values.}
\label{correlation}
\end{figure}

In order to describe the correlation of sub-channels, in this paper we utilise the spatial correlation function, expressed as
\begin{equation}\label{STCF} 
{R_h}(\Delta {x_T},{\kern 1pt} {\kern 1pt} {\kern 1pt} \Delta {x_R}) = E\{ {h_{p,q}}{\kern 1pt} (t)h_{p'q'}^ * (t)\}
\end{equation}
where $\Delta {x_R}$ and $\Delta {x_T}$ are the antenna spacings at the Rx and Tx respectively.
In Fig. \ref{correlation}, the spatial correlation function of the Tx is shown. From these two figures, it can be observed that in this massive MIMO channel model, a higher cluster evolution factors can result in a smaller number of clusters shared by different antennas. This phenomenon will lead to lower correlation between sub-channels. Compared with conventional MIMO and i.i.d. MIMO channel models, the fading characteristic of this massive MIMO channel model used in this paper is significantly different, indicating the necessity to use specific massive MIMO channel models to design and test the performance of such systems.

\section{theoretical analysis}\label{S3}
\subsection{The theoretical BER expression of TDMA-MU-SM system}
For a TDMA-MU-SM system, at each time instant, all Tx antennas only transmit data to one user. This is similar to the conventional point-to-point scenario and its theoretical BER can be approached via the union bound method mentioned in \cite{YuThesis,RenzoTVT12SMBER, Hadayat05SMBER,BER1}. Here, we can define ${{x_{{n_t},l}}}$ as the transmitted symbol $l$ from antenna $n_t$, $l = 1,2...M,{n_t} = 1,2...{N_t}$.
The theoretical  upper-bound expression of the BER for the $k$-th user can be expressed as
\begin{align}\label{SMBER1} 
\rm{ABER_k} \leqslant &\frac{1}{{{N_t}M}}\frac{1}{m}\sum\limits_{{n_t} \ne {{\tilde n}_t}} \sum\limits_{l \ne \tilde l}\nonumber \\ &{\left[ {{N_H}({x_{{n_t},l}} \to {x_{{{\tilde n}_t},\tilde l}})P({x_{{n_t},l}} \to {x_{{{\tilde n}_t},\tilde l}}\left| H_k \right.)} \right]}
\end{align}
where ${{N_H}({x_{{n_t},l}} \to {x_{{{\tilde n}_t},\tilde l}})}$ is the Hamming distance between ${{x_{{n_t},l}}}$ and ${x_{{{\tilde n}_t},\tilde l}}$. The pairwise error probability (PEP) that ${{x_{{n_t},l}}}$ is detected to ${x_{{{\tilde n}_t},\tilde l}}$ under the channel $H_k$ is expressed as ${P({x_{{n_t},l}} \to {x_{{{\tilde n}_t},\tilde l}}\left| H_k \right.)}$. 
For a ML detector, as it detects the sample based on the Euclidean distance, the error only happens when ${\left\| y_k - H_k {x_{{{n}_t}, l}} \right\|^2}>{\left\| y_k -H_k{x_{{{\tilde n}_t},\tilde l}} \right\|^2}$.  If we define $e = {x_{{n_t},l}}-{x_{{{\tilde n}_t},\tilde l}}$, the PEP can be expressed as \cite{Hadayat05SMBER}
\begin{align}\label{BER9-1} 
{P({x_{{n_t},l}} \hspace{-0.06cm}\to\hspace{-0.06cm} {x_{{{\tilde n}_t},\tilde l}}\left| H_k \right.)}&\hspace{-0.06cm}=\hspace{-0.06cm} P({\left\| y\hspace{-0.06cm}-\hspace{-0.06cm}H_k {x_{{n_t}, l}} \right\|^2}\hspace{-0.06cm}>\hspace{-0.06cm}{\left\| y \hspace{-0.06cm}-\hspace{-0.06cm}H_k{x_{{{\tilde n}_t},\tilde l}} \right\|^2}\hspace{-0.06cm})\nonumber \\ &= Q(\gamma  {\left\| {H_k e } \right\|^2})
\end{align}
where $\gamma$ is the signal-to-noise ratio (SNR) and ${Q(.)}$ is the Q-function. Based on the alternative integral expression of Q-function using MGF-based approach \cite{Hadayat05SMBER}, \cite{BER1}, (\ref{BER9-1}) can be rewritten as
 \begin{align}\label{BER9-2} 
{P({x_{{n_t},l}} \to {x_{{{\tilde n}_t},\tilde l}}\left| H_k \right.)} = \frac{1}{\pi }\int_0^{\pi /2} {{M}(\gamma\frac{1}{{{{\sin }^2}\theta }})} d\theta .
\end{align}
In (\ref{BER9-2}), ${M}(.)$ is the MGF of ${\left\| {H_k e } \right\|^2}$, which can be expressed as
\begin{equation}\label{BER11} 
M(s) = \prod\limits_{j = 1}^{{N_r}} {{{\left( {1 - s{\lambda _i}\mu } \right)}^{ - 1}}}
\end{equation}
where $\lambda _i$ is the eigenvalues of $R_R$ and $\mu$ is the eigenvalue of $e{e^H}{R_T}$. Substituting (\ref{BER9-2}) and (\ref{BER11}) into (\ref{SMBER1}),  the final expression of the theoretical BER expression for the $k$-th user of a TDMA-MU-SM is 
\begin{align}\label{BER_tdsm_f} 
&\rm{ABER_k} \leqslant \frac{1}{\pi }\frac{1}{{{N_t}M}}\frac{1}{m} \nonumber \\ &\sum\limits_{{n_t} \ne {{\tilde n}_t}} {\sum\limits_{l \ne \tilde l} {\left[ {{N_H}({x_{{n_t},l}} \hspace{-0.06cm}\to\hspace{-0.06cm} {x_{{{\tilde n}_t},\tilde l}})\int_0^{\frac{\pi }{2}} {\prod\limits_{j = 1}^{{N_r}} {{{(1 \hspace{-0.06cm}- \hspace{-0.06cm}\frac{\gamma }{{{{\sin }^2}\theta }}{\lambda _i}\mu )}^{ - 1}}d\theta } } } \right]} }.
\end{align}

\subsection{Theoretical BER expression of BD-MU-SM system}
For the BD-MU-SM system, we can follow the principle described above to derive the theoretical BER expression. However, due to the fact that some data are mapped to the beamforming pattern index, the error case should be expressed as ${x_{{n_b},l}} \to {x_{{n_{\tilde b}},\tilde l}}$ , which means the transmitted symbol $l$ using the beamforming pattern $n_b$ is detected to symbol $\tilde l$ carried by the $n_{\tilde b}$ beamforming pattern, $(l, \tilde l)= 1,2...M;  (n_b, n_{\tilde b})= 1,2...N$.
The theoretical  upper-bound expression of the BER for the $k$-th user can be expressed as
\begin{align}\label{BER_BDSM} 
&\rm {ABEP_k} \leqslant \frac{1}{{NM}}\frac{1}{m} \nonumber \\ &\sum\limits_{{n_b} \ne {{\tilde n}_t}} {\sum\limits_{l \ne \tilde l} {\left[ {{N_H}({x_{{n_b},l}} \hspace{-0.06cm}\to\hspace{-0.06cm} {x_{{n_{\tilde b}},\tilde l}})P({x_{{n_b},l}}\hspace{-0.06cm} \to\hspace{-0.06cm} {x_{{n_{\tilde b}},\tilde l}}\left| {{H_E}^k} \right.)} \right]}}
\end{align}
where ${N_H}({x_{{n_b},l}} \to {x_{{n_{\tilde b}},\tilde l}})$ is the Hamming distance between ${x_{{n_b},l}}$ and ${x_{{n_{\tilde b}},\tilde l}}$,  $P({x_{{n_b},l}} \to {x_{{n_{\tilde b}},\tilde l}}\left| {{H_E}^k} \right.)$ is the PEP of the error that occurs under the effective channel ${H_E}$. From (\ref{CodedMat}),  we can write the expression of ${H_E}^k$ as
\begin{equation}\label{EFFh} 
{H_E}^k = {H_k}{{\tilde V}_k} = R_{R,k}^{_2^1}{H_W}R_{T,k}^{_2^1}{{\tilde V}_k} = R_{R,k}^{_2^1}{H_W}{({R_{T,k}}{{\tilde V}^2})^{\frac{1}{2}}}.
\end{equation}
Thus, for the effective channel, the Rx correlation matrix is not changed, but the Tx correlation matrix is changed by the precoding matrix. The Tx effective correlation matrix is defined as ${R_{E,T,k}} = {R_{T,k}}{{\tilde V}^2}$. 
For the ML detector of the BD-MU-SM system, the error will occur for the case that ${\left\| y_k -{H_E}^k{\rho _k} {x_{{n_b},l}} \right\|^2} > {\left\| y_k -{H_E}^k{\rho _k} {x_{{n_{\tilde b}},\tilde l}}  \right\|^2}$.  We can also define the error vector $e={x_{{n_b},l}}- {x_{{n_{\tilde b}},\tilde l}}$.
The expression of PEP in (\ref{BER_BDSM}) can be expressed as
\begin{align}\label{BDPEP} 
&P({x_{{n_b},l}}\hspace{-0.06cm}\to \hspace{-0.06cm}{x_{{n_{\tilde b}},\tilde l}}\left| {{H_E}^k} \right.)\nonumber \\ &\hspace{-0.06cm}=\hspace{-0.06cm} P({\left\| y_k \hspace{-0.06cm}-\hspace{-0.06cm}{H_E}^k{\rho _k} {x_{{n_b},l}} \right\|^2}\hspace{-0.06cm} > \hspace{-0.06cm}{\left\| y_k \hspace{-0.06cm}-\hspace{-0.06cm}{H_E}^k{\rho _k} {x_{{n_{\tilde b}},\tilde l}}  \right\|^2})\hspace{-0.06cm}\nonumber \\ &=\hspace{-0.06cm} Q(\gamma_E  {\left\| {{H_E}^k e } \right\|^2})
\end{align}
where $\gamma_E$ is the effective SNR, which can be expressed as $\gamma_E=\rho _k^2\gamma$. Similar to the TDMA-MU-SM case, the PEP can be expressed by the MGF-based approach in (\ref{BER9-2}) using the MGF of $\left\| {H_E}^k e \right\|^2$. However, for the BD case, the value of $\mu$ in (\ref{BER11}) is the eigenvalue of $e{e^H}{R_{E,T,k}}$. Finally, we can express the final theoretical BER expression for the $k$-th user of a BD-MU-SM as
\begin{align}\label{BER_BDF} 
&\rm{ABER_k} \leqslant \frac{1}{\pi }\frac{1}{{NM}}\frac{1}{m} \nonumber \\ &\sum\limits_{{n_b} \ne {{\tilde n}_b}} {\sum\limits_{l \ne \tilde l} {\left[ {{N_H}({x_{{n_b},l}} \hspace{-0.06cm}\to\hspace{-0.06cm} {x_{{{\tilde n}_b},\tilde l}})\int_0^{\frac{\pi }{2}} {\prod\limits_{j = 1}^{{N_r}} {{{(1\hspace{-0.06cm} -\hspace{-0.06cm} \frac{{{\gamma _E}}}{{{{\sin }^2}\theta }}{\lambda _i}\mu )}^{ - 1}}d\theta } } } \right]} }.
\end{align}

\vspace{0.5cm}
\section{Results and analysis}\label{S4}
In this section, different simulation results are presented to enable evaluation of the performance of SM systems for a massive channel model. For all simulations, the base station is assumed to be equipped with 64 transmit antennas with 2 receive antennas employed for each user. To ensure a fair comparison, we assume that the transmit power and the data rate of each user are the same with all simulation results obtained through Monte Carlo simulation. For each point in figures that follow the results of 10 runs (where each run transmits 100000 symbols) are averaged to obtain the final value.

\begin{figure}[t]
 \hspace{-0.2cm} \includegraphics[width=3.5 in]{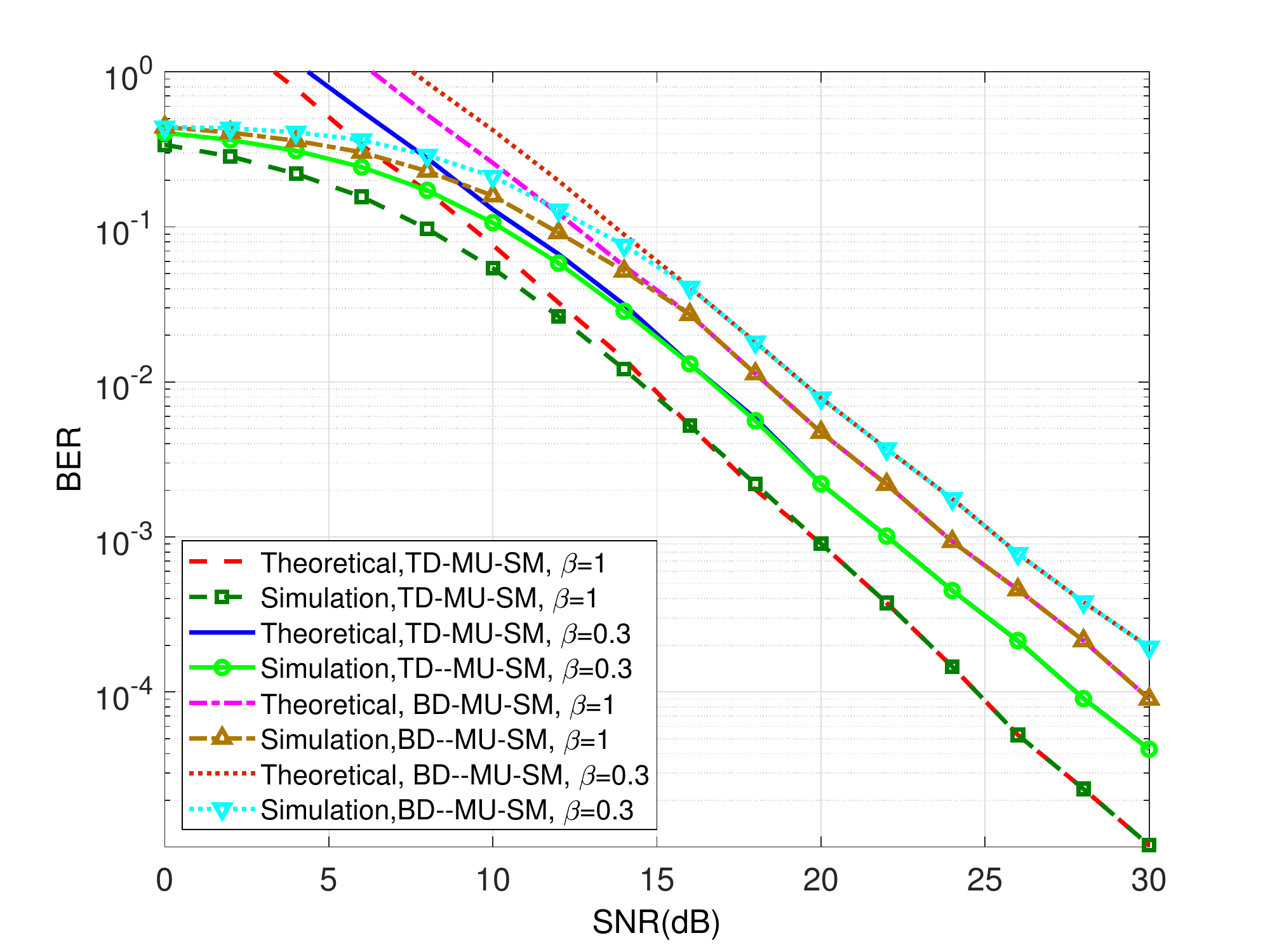}
\caption{Theoretical and simulated BER results of SM systems with massive MIMO KBSM.}
\label{theoreticalFig}
%\label{fig_sim}
\end{figure}

In Fig. \ref{theoreticalFig}, theoretical results of the BER performance for a TDMA-MU-SM and a BD-MU-SM system assuming the massive channel model, are compared with each of the simulation results assuming different clusters evolution factors, $\beta $ with the data rate set at $m$=7 bits/symbol/user for both systems. For the TDMA-SM system, 6 bits are carried by the antenna indices, and a BPSK modulation scheme is employed to carry the other bits. For the BD-SM system, 32 beamforming patterns are used by each user, thus the index can encode 5 bits of information, the other 2 bits encoded in QPSK modulation symbols. From this figure, we can observe the following: First, it shows an excellent agreement between the approximation of the theoretical derivation by the simulation results for all systems under different channel configurations, particularly in the high SNR regime, ie where SNR $\ge 15$~dB. Based on this, we can conclude that the theoretical BER upper-bound offers reasonable accuracy, providing reasonable assurance in the theoretical derivation; Second, it is clear that for the same data rate and using the same transmit power for each user, the TDMA-MU-SM system offers better performance in terms of the BER than the BD-MU-SM system even when the BD system has the CSI at the Tx. This is because for TDMA-MU-SM systems, at each time instant only one antenna transmits signals to a single user, thus IUI is avoided. For the BD-MU-SM system, as the precoding matrix is applied to the transmitted signal, this results in a loss of transmit power for IUI cancellation. Consequently, the effective power for the signal transmission in the BD scheme is lower. However, it should be pointed out that the total system data rate for the BD-MU-SM system is higher than the TDMA system (${K \times m}~ \rm{Vs.}~{m}$) at the expense of using more transmit power. Thus for the TDMA scheme, in order to obtain the same data rate, it would require the use of higher order modulation schemes. From this analysis, we observe that the TDMA scheme is suitable for use in light-load MU systems that require low system complexity and  high accuracy, while the BD scheme is suitable for heavy-load systems that require high system capacity; Third, we note that in channel conditions with a higher cluster evolution factor $\beta$, both the TDMA-MU-SM and BD-MU-SM systems obtain better BER performance when compared with that of lower cluster evolution factors. This phenomena can be explained using the insight offered from Section III. In Fig. \ref{ClusterShare} note that with higher cluster evolution factors, the probability that different antennas observe different sets of clusters is increased,  resulting in lower correlation between the different sub-channels as shown in Fig. \ref{correlation}. In the TDMA scheme, lower correlation between sub-channels make it easier for the Rx to correctly detect the active antennas results in the system BER performance being better as the TDMA-SM scheme can be treated as a conventional SM system. While for the BD scheme, the lower correlation between sub-channels decreases the difficulty in eliminating the IUI, which is represented by the lower power assumption of the precoding matrix, results in more power being used to transmit effective data and the BER performance is improved.

\begin{figure}[t]
 \hspace{-0.2cm} \includegraphics[width=3.5 in]{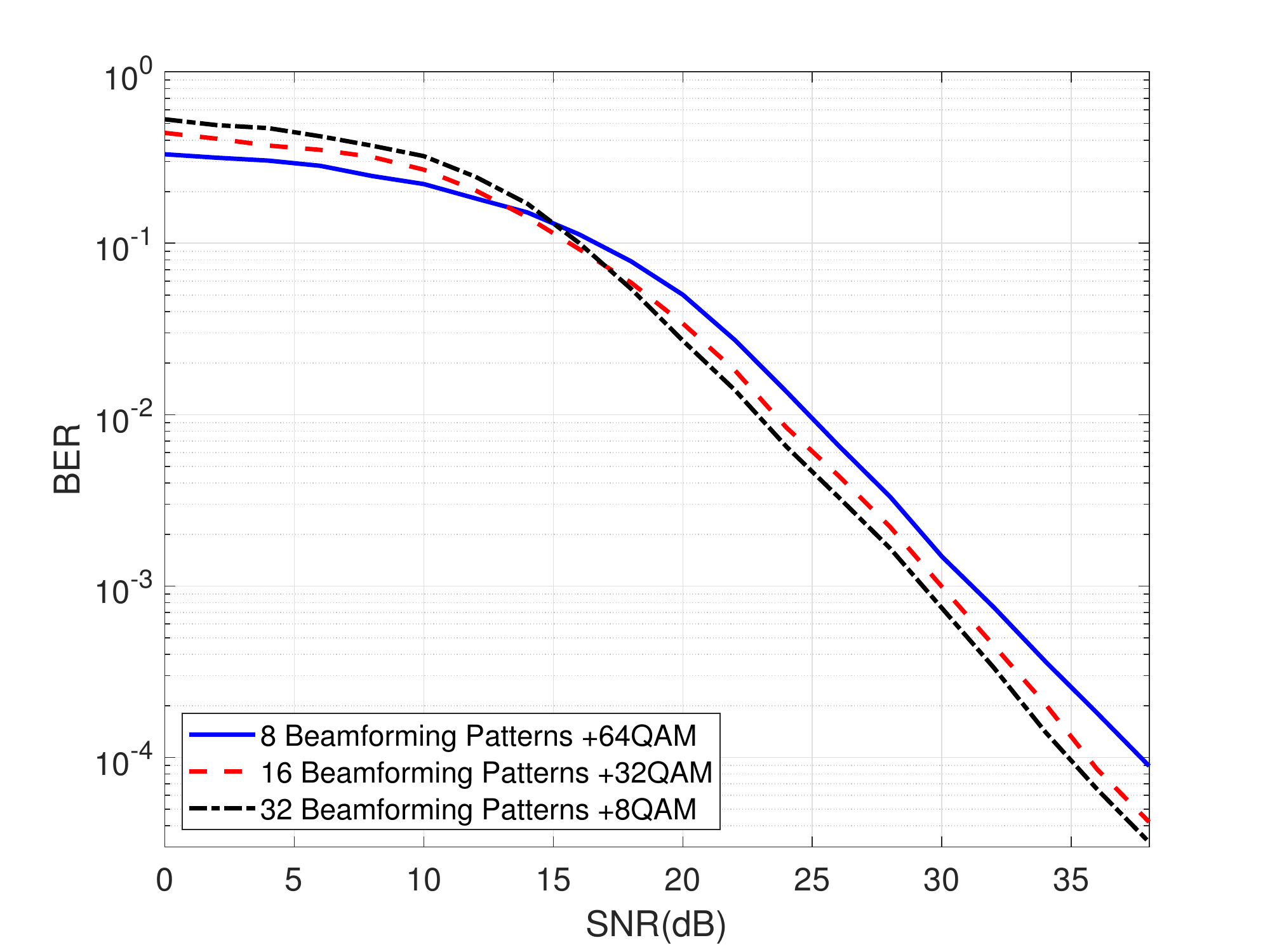}
\caption{BER performance of BD-MU-SM systems using different system settings, $\beta=1$.}
\label{Beam1}
%\label{fig_sim}
\end{figure}

\begin{figure}[t]
 \hspace{-0.2cm} \includegraphics[width=3.5 in]{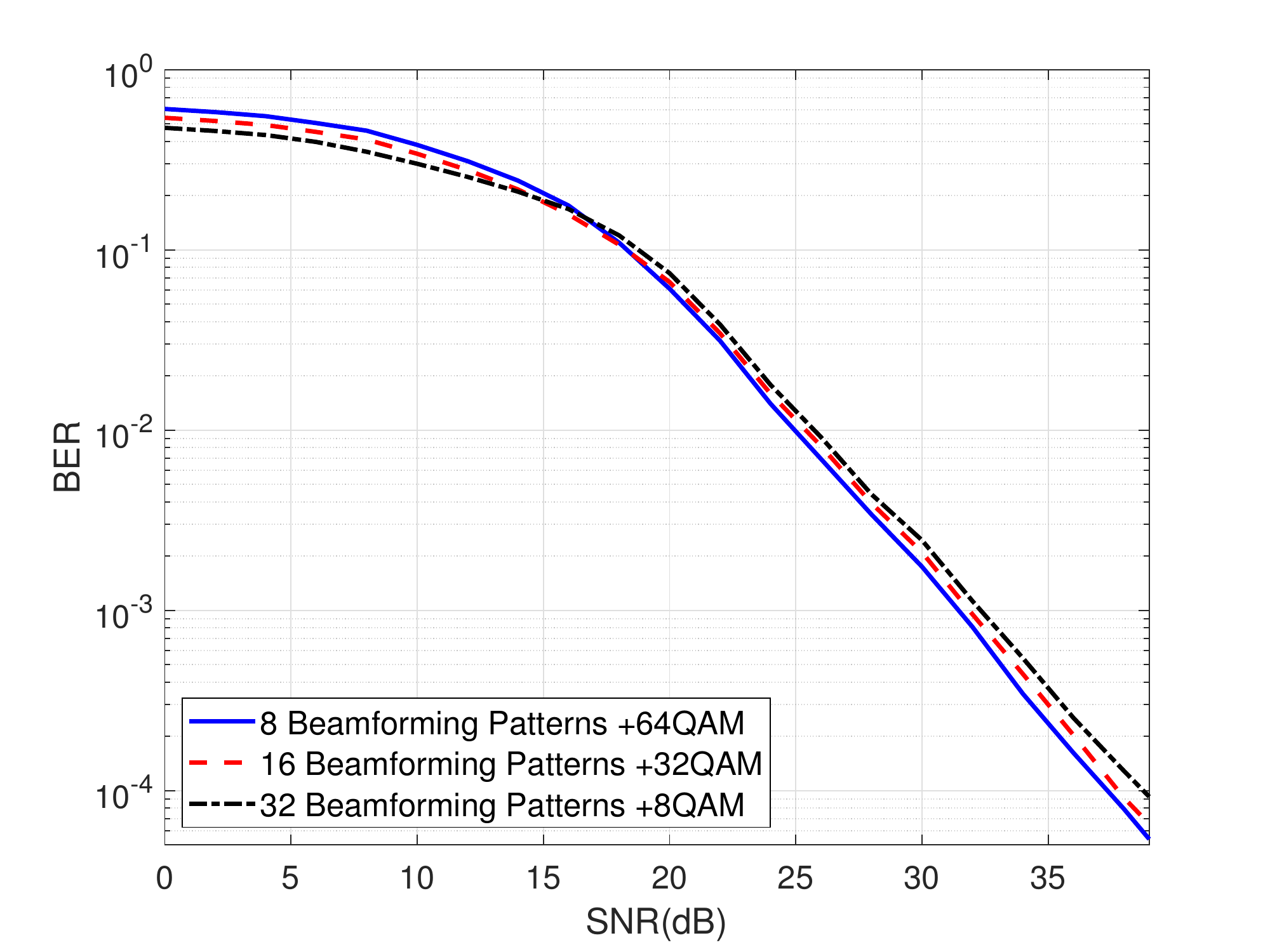}
\caption{BER performance of BD-MU-SM systems using different system settings, $\beta=0.3$. }
\label{Beam2}
%\label{fig_sim}
\end{figure}
% that's all folks
% that's all folks

In  Figs. \ref{Beam1} and \ref{Beam2}, the BER performance of the BD-MU-SM system is investigated for different systems settings and different channel assumptions ($\beta=0.3$ and $1$) via simulation results. It is assumed that there are 8 users and that the data rate of each user is $m$=9 bits/symbol. Comparing these two figures, we observe that higher cluster evolution factors results in a better BER performance. In Fig. \ref{Beam1}, we note that in the low SNR regime, using fewer beamforming patterns to carry information results in better BER performance. While in the high SNR regime, using more beamforming patterns plus lower order modulation schemes can achieve a better BER performance. The crossover occurs at approximately $15$~dB. In respect of these phenomena, we offer the following explanations. When using the BD scheme, a smaller number of beamforming patterns results in the power being more concentrated. As the number of patterns is small, the task of the detector is easier thus decreasing the probability of errors occurring. But in the high SNR regime, the benefit of lower index detection error is outstripped by the use of lower order modulation schemes. Thus, in the high SNR regime, using larger numbers of beamforming patterns plus lower modulation schemes offers better BER performance. If we consider Fig. \ref{Beam2}, we reach a similar conclusion. However, compared with the case for higher cluster evolution factors, the difference between the different system settings is not so significant. The position of the crossover is located just below $20$~dB. The reason for this phenomenon is the high correlation between the sub-channels. Due to the high correlation, the benefit from using beamforming patterns to carry information is reduced. In this case, the primary determination of system performance is the SNR, thus the different system settings offer no significant performance advantage. 

\begin{figure}[t]
 \hspace{-0.2cm} \includegraphics[width=3.5 in]{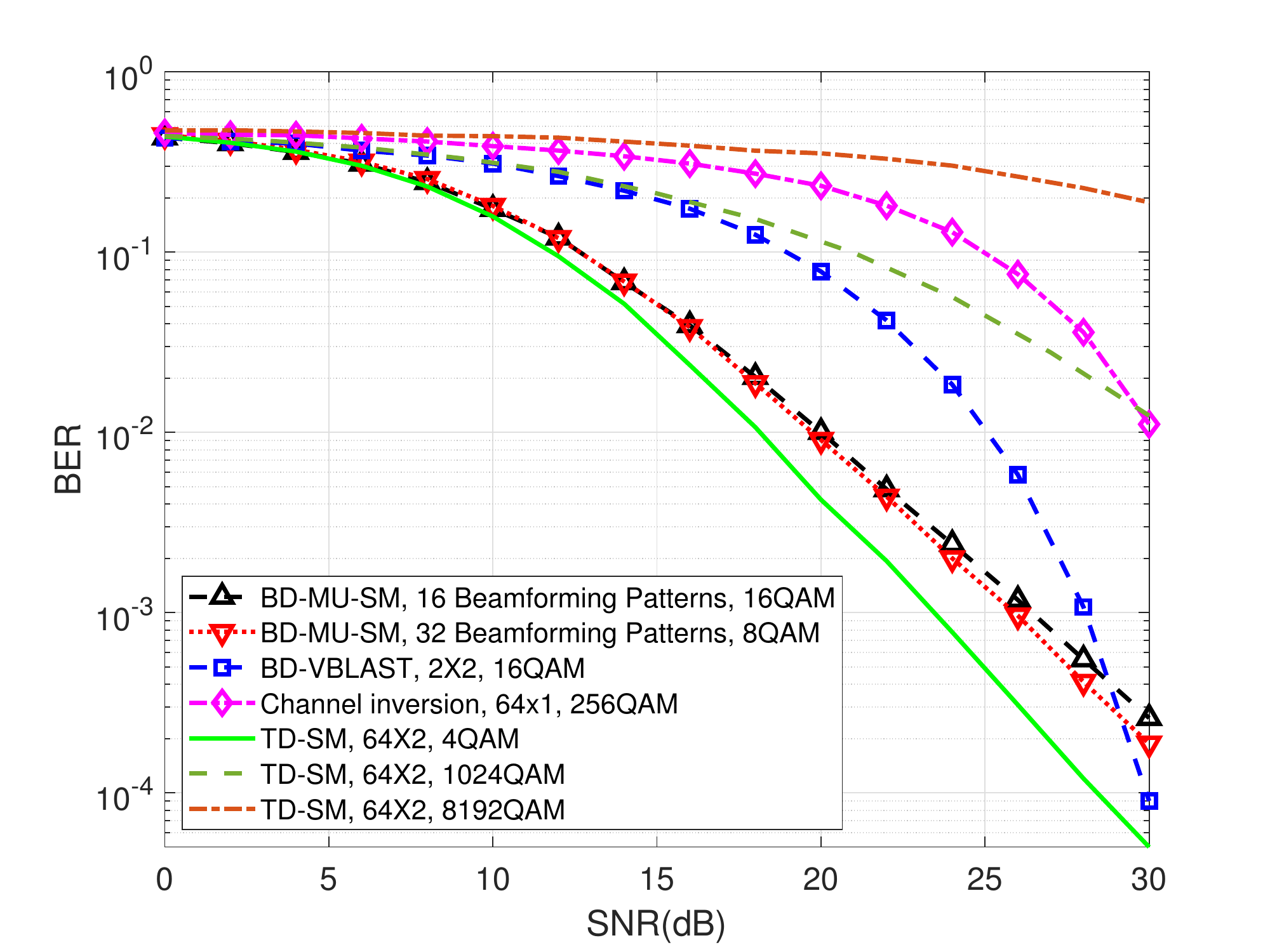}
\caption{BER performance of different systems with 4 users, $\beta=0.3$, data rate= 8 bits/symbol/user.}
\label{MU1}
%\label{fig_sim}
\end{figure}

\begin{figure}[t]
 \hspace{-0.2cm} \includegraphics[width=3.5 in]{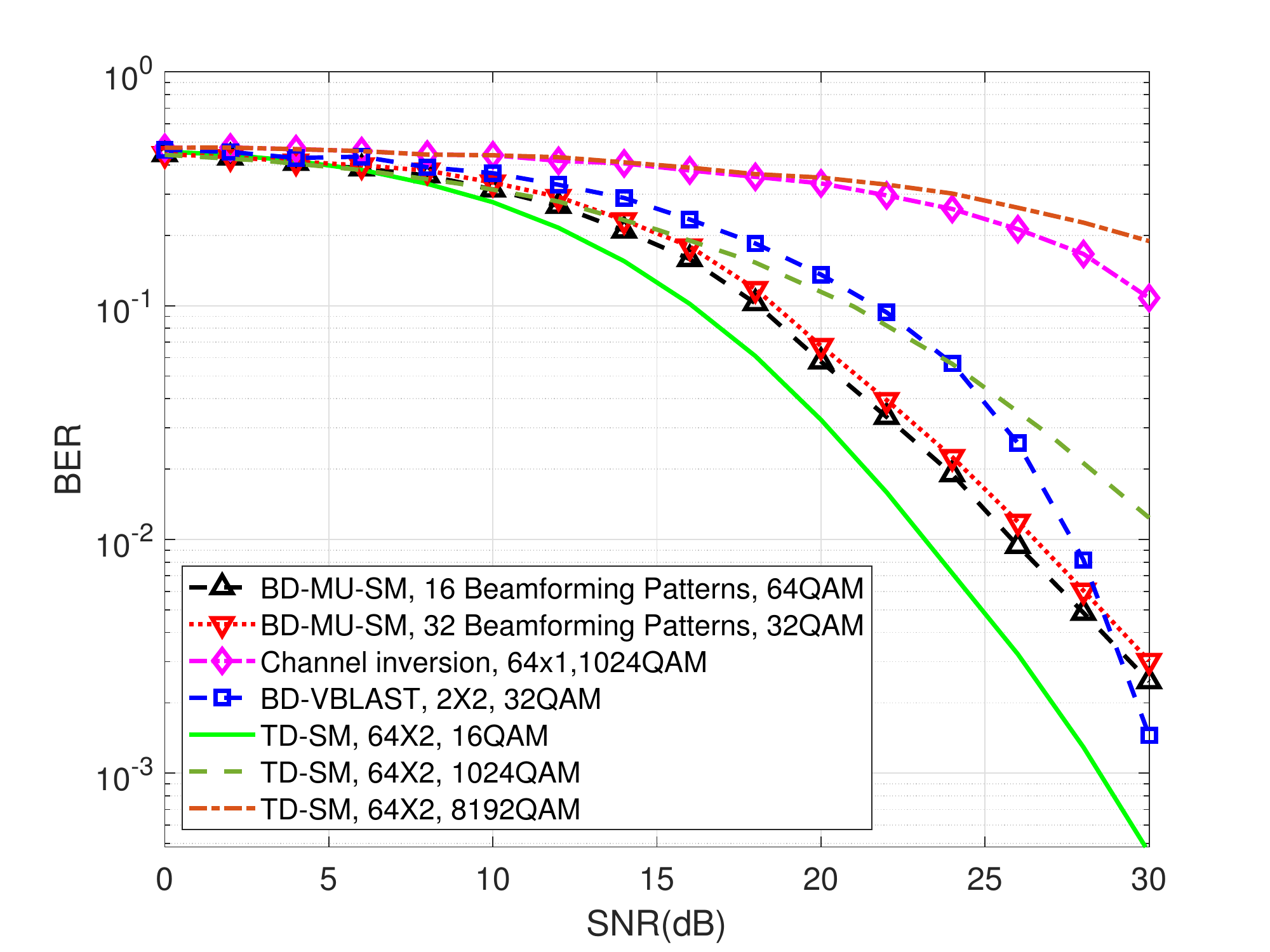}
\caption{BER performance of different systems with 4 users, $\beta=0.3$, data rate= 10 bits/symbol/user.}
\label{MU2}
%\label{fig_sim}
\end{figure}

\begin{figure}[t]
 \hspace{-0.2cm} 
 \includegraphics[width=3.5 in]{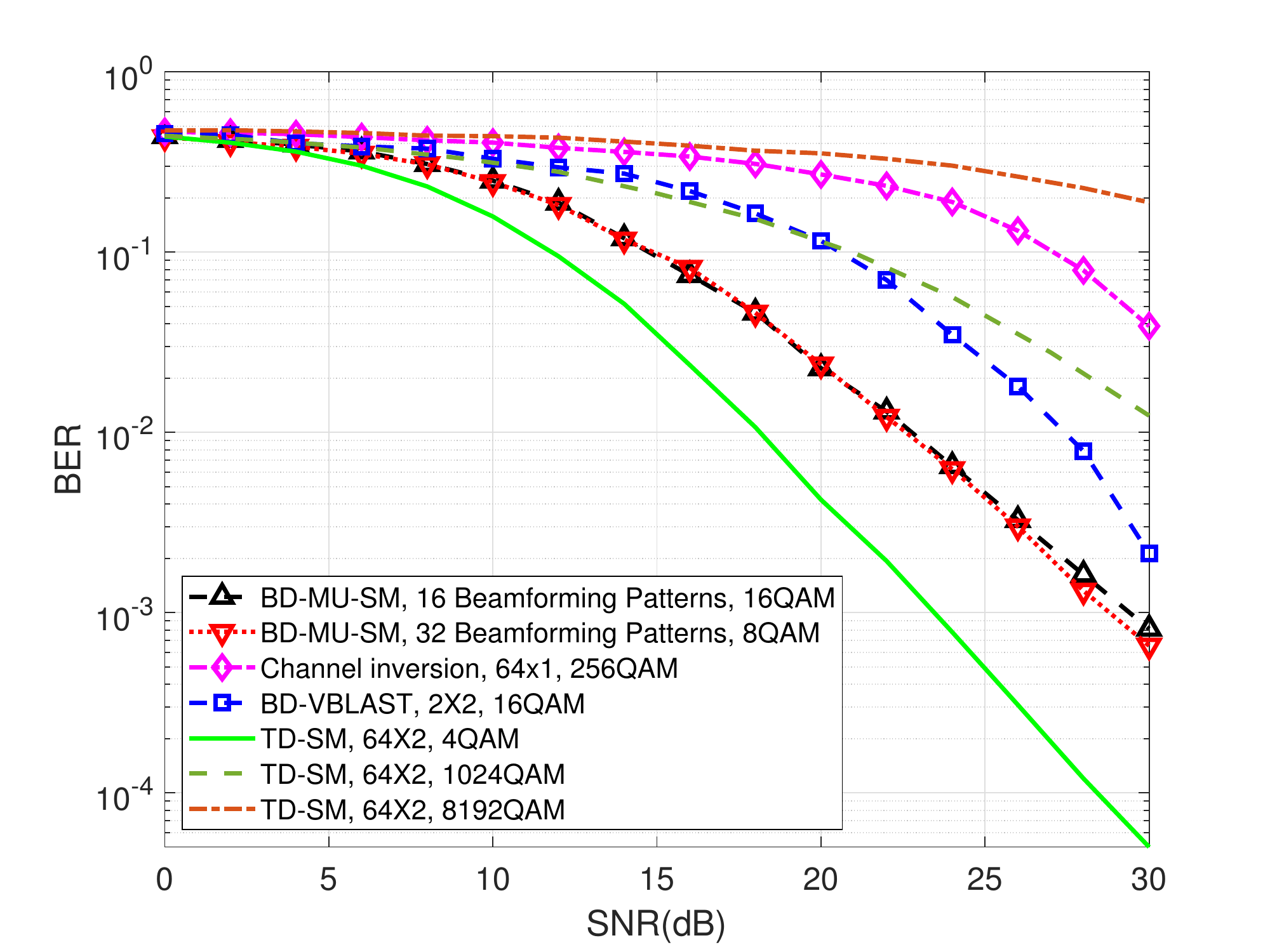}
\caption{BER performance of different systems with 16 users, $\beta=0.3$, data rate= 8 bits/symbol/user.}
\label{MU3}
%\label{fig_sim}
\end{figure}

\begin{figure}[t]
 \hspace{-0.2cm} \includegraphics[width=3.5 in]{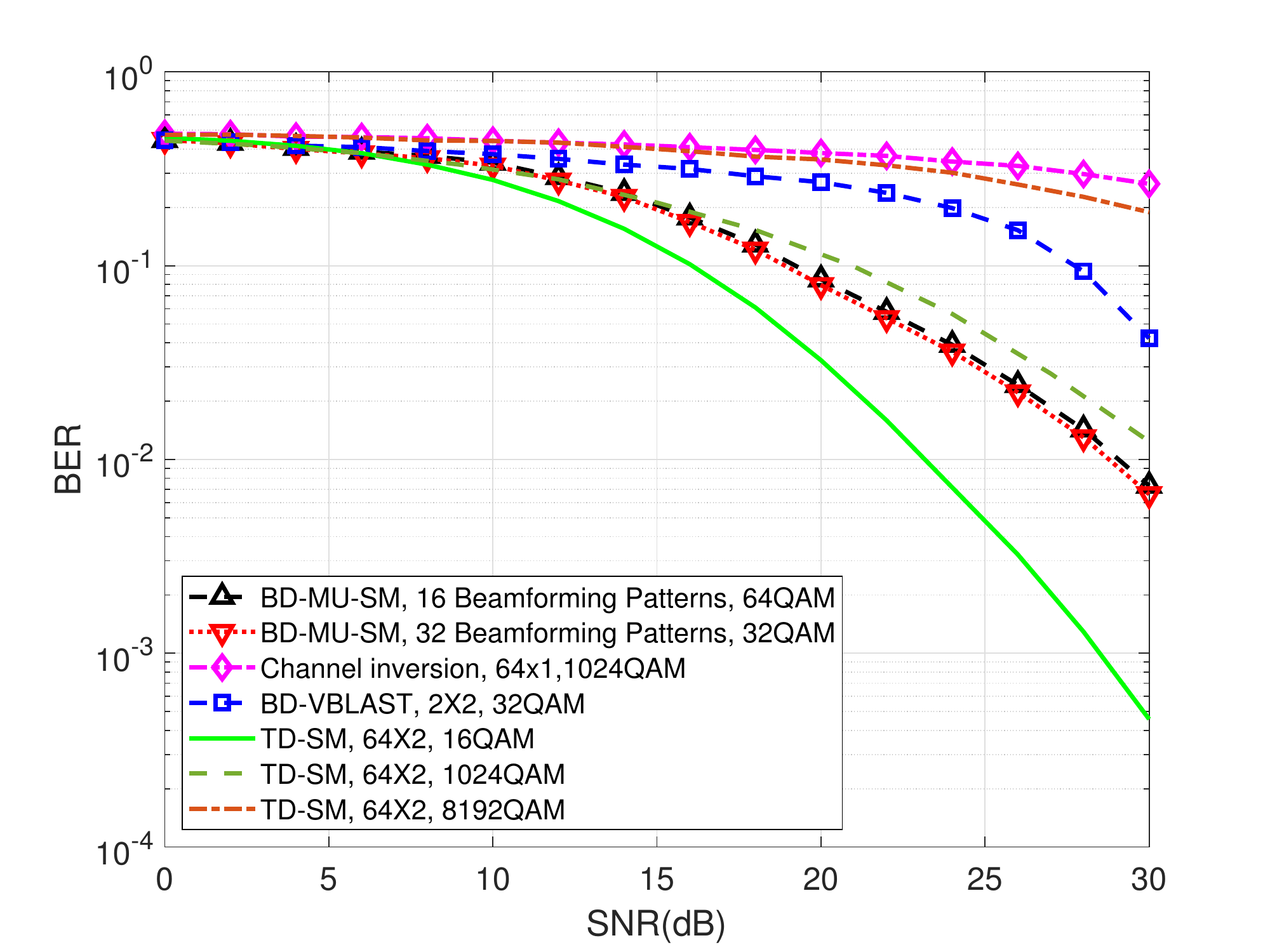}
\caption{BER performance of different systems with 16 users, $\beta=0.3$, data rate= 10 bits/symbol/user.}
\label{MU4}
%\label{fig_sim}
\end{figure}

In Figs. \ref{MU1}--\ref{MU4}, the simulated BER performance of the TDMA-MU-SM system and the BD MU-SM system are compared with the V-BLAST approach and the conventional channel inversion method under the massive channel model with different data rates and numbers of users. In all simulations, the evolution factor for the channel is set at $\beta  =0.3$. For the V-BLAST system, in order to make the comparison fair, the BD precoding scheme is used at the Tx side to eliminate inter-user interference and the ML detector is employed as the Rx algorithm. As the V-BLAST system requires that the receive antenna number should not be smaller than the number of layers, the number of the beamforming patterns for the V-BLAST system is configured as two for each user. For the channel inversion method, the precoding matrix is given as the Moore-Penrose pseudo inverse of the channel matrix $H$, which can be expressed as $W = {H^H}{(H{H^H})^{ - 1}}$. In this system, only one receive antenna is available to each user.

% that's all folks

% that's all folks

% that's all folks

% that's all folk

First from all of the figures it is clear that for all of the systems, when working at lower data rates, that the system performance is better as the power averaged to each bit is higher. Second, for both BD-SM and BD-V-BLAST systems, when the number of users increases, the system performance degrades. This is because with more users, the system requires more power to eliminate the inter-user-interference. As the transmit power for the system is fixed, the power available for transmitting effective signals will decrease. So the system performance will degrade. For the TDMA-SM system, as only one user is being served, the BER performance is not affected by the number of users. For a fair and comprehensive comparison, in addition to the modulation order that can obtain the same data rate of one user, we have also used much higher  modulation schemes (1024 QAM,system data rate= 16 bits/symbol and 8192 QAM, system data rate= 19 bits/symbol ) to increase the total data rate of the system. When the data rate for each user is the same, compared to the previously discussed two BD-based methods, the TDMA-SM scheme can obtains the best performances for all cases. But for the TDMA-SM scheme, the system's total data rate is much lower than the other systems under heavy-load scenarios as it can only serve one user in each time slot. Even when some very high-order modulation schemes are used, the TDMA-SM system can still not achieve a similar total system data rate. At the same time, its BER performance is poor, which indicate that the TDMA-SM system is not suitable for a MU system with lots of users and requires high system data rate. But we cannot neglect the advantage of the TDMA-SM system as it can provide an efficient solution to build low-cost massive MIMO systems. For a low data rate system which requires high accuracy and low system complexity, the TDMA-SM is still a promising solution. Compared to the BD-SM systems and BD-V-BLAST systems, the BD-SM system offers a greater enhancement of the BER performance in the SNR range of $< 28$~dB. For the V-BLAST system, in the case of a low data rate and a small numbers of users, it offers a better performance in the extremely high SNR range (SNR  $> 28$~dB). However, for the case of high data rates with a large numbers of users, the BD-SM system is superior to the BD-V-BLAST system. Compared with the conventional channel inversion method, the advantage of the BD base precoding scheme is clear. It benefits from supporting multiple receive antennas, and so the BER performance of all BD based system is superior because of the receive diversity gain. In Table 1, some detailed simulation results are also provided.   

\begin{table*}[t]
\label{BERform}
\centering
  \caption{BER performance of different systems.}
\begin{tabular}{|l|l|l|l|l|l|l|l|l|l|l|l|l|}
\hline
Data rate                                                                    & \multicolumn{6}{c|}{\textbf{8 bits/symbol/user}}                                                                                          & \multicolumn{6}{c|}{\textbf{10 bits/symbol/user}}                                                                                         \\ \hline
Users                                                                        & \multicolumn{3}{c|}{\textbf{4}}                                     & \multicolumn{3}{c|}{\textbf{16}}                                    & \multicolumn{3}{c|}{\textbf{4}}                                     & \multicolumn{3}{c|}{\textbf{16}}                                    \\ \hline
SNR (dB)                                                                     & \multicolumn{1}{c|}{\textbf{10}} & \multicolumn{2}{c|}{\textbf{30}} & \multicolumn{1}{c|}{\textbf{10}} & \multicolumn{2}{c|}{\textbf{30}} & \multicolumn{1}{c|}{\textbf{10}} & \multicolumn{2}{c|}{\textbf{30}} & \multicolumn{1}{c|}{\textbf{10}} & \multicolumn{2}{c|}{\textbf{30}} \\ \hline
\begin{tabular}[c]{@{}l@{}}BD-MU-SM, \\ 16 Beamforming Patterns\end{tabular} & 0.17369                          & \multicolumn{2}{l|}{0.00026}     & 0.31412                          & \multicolumn{2}{l|}{0.00247}     & 0.24917                          & \multicolumn{2}{l|}{0.00081}     & 0.33983                          & \multicolumn{2}{l|}{0.00732}     \\ \hline
\begin{tabular}[c]{@{}l@{}}BD-MU-SM,\\ 32 Beamforming Patterns\end{tabular}  & 0.18058                          & \multicolumn{2}{l|}{0.00019}     & 0.33625                          & \multicolumn{2}{l|}{0.00303}     & 0.24456                          & \multicolumn{2}{l|}{0.00065}     & 0.32812                          & \multicolumn{2}{l|}{0.00066}     \\ \hline
BD-VBLAST                                                                    & 0.31014                          & \multicolumn{2}{l|}{0.00009}     & 0.36895                          & \multicolumn{2}{l|}{0.00247}     & 0.32982                          & \multicolumn{2}{l|}{0.00213}     & 0.37639                          & \multicolumn{2}{l|}{0.04221}     \\ \hline
Channel inversion                                                            & 0.38751                          & \multicolumn{2}{l|}{0.01109}     & 0.44012                          & \multicolumn{2}{l|}{0.00145}     & 0.30817                          & \multicolumn{2}{l|}{0.03886}     & 0.44113                          & \multicolumn{2}{l|}{0.26444}     \\ \hline
TD-SM, 4 QAM                                                                  & 0.15768                          & \multicolumn{2}{l|}{0.00005}     & 0.15768                          & \multicolumn{2}{l|}{0.00005}     & 0.15768                          & \multicolumn{2}{l|}{0.00005}     & 0.15768                          & \multicolumn{2}{l|}{0.00005}     \\ \hline
TD-SM, 1024 QAM                                                              & 0.30894                          & \multicolumn{2}{l|}{0.01238}     & 0.30894                          & \multicolumn{2}{l|}{0.01238}     & 0.30894                          & \multicolumn{2}{l|}{0.01238}     & 0.30894                          & \multicolumn{2}{l|}{0.01238}     \\ \hline
TD-SM, 8192 QAM                                                              & 0.44012                          & \multicolumn{2}{l|}{0.18896}     & 0.44012                          & \multicolumn{2}{l|}{0.18896}     & 0.44012                          & \multicolumn{2}{l|}{0.18896}     & 0.44012                          & \multicolumn{2}{l|}{0.18896}     \\ \hline
\end{tabular}
\end{table*}

\section{Conclusions}\label{S5}
In this paper, we have investigated the performance of SM systems when operating in a massive MIMO KBSM for a range of different system settings and channel configurations. We observe that higher evolution factors of the scatters results in a better BER performance for SM systems because of the lower correlation among the sub-channels. The TDMA-MU-SM system can offer a better BER performance at the cost of a lower system data rate. The BD-MU-SM system provides the best trade-off between system data rate and BER performance, and as such is a better option for heavy-load MU wireless communication systems. Compared to the BD-V-BLAST and conventional channel inversion systems, the two proposed SM systems offer a better BER performance for a variety of data rates and numbers of users, particularly in the low SNR regime. All of the numerical and simulation results make clear the advantage that arises from using SM for massive MIMO systems in terms of system performance, complexity and flexibility.

\begin{IEEEbiography}
[{\includegraphics[width=1in,height=1.25in,clip,keepaspectratio]{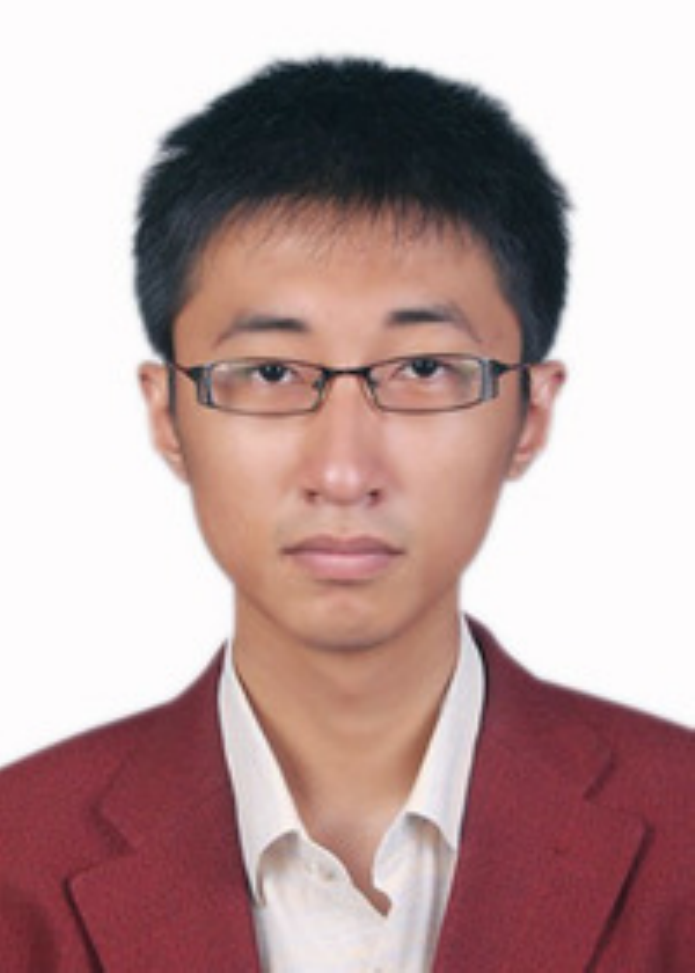}}]
{Yu Fu} received his BSc degree in Computer Science
from Huaqiao University, Fujian, China, in 2009,
the MSc degree in Information Technology (Mobile
Communications) from Heriot-Watt University, Edinburgh,
U.K., in 2010, and the PhD degree in Wireless Communications from Heriot-Watt University, Edinburgh, UK, in 2015. He has been a Postdoc Research Associate of Heriot-Watt University since 2015. His main research interests include advanced MIMO communication technologies, wireless channel modeling and simulation, RF tests, and software defined networks. Email: y.fu@hw.ac.uk.
\end{IEEEbiography} 

\begin{IEEEbiography}[{\includegraphics[width=1in,height=1.25in,clip,keepaspectratio]{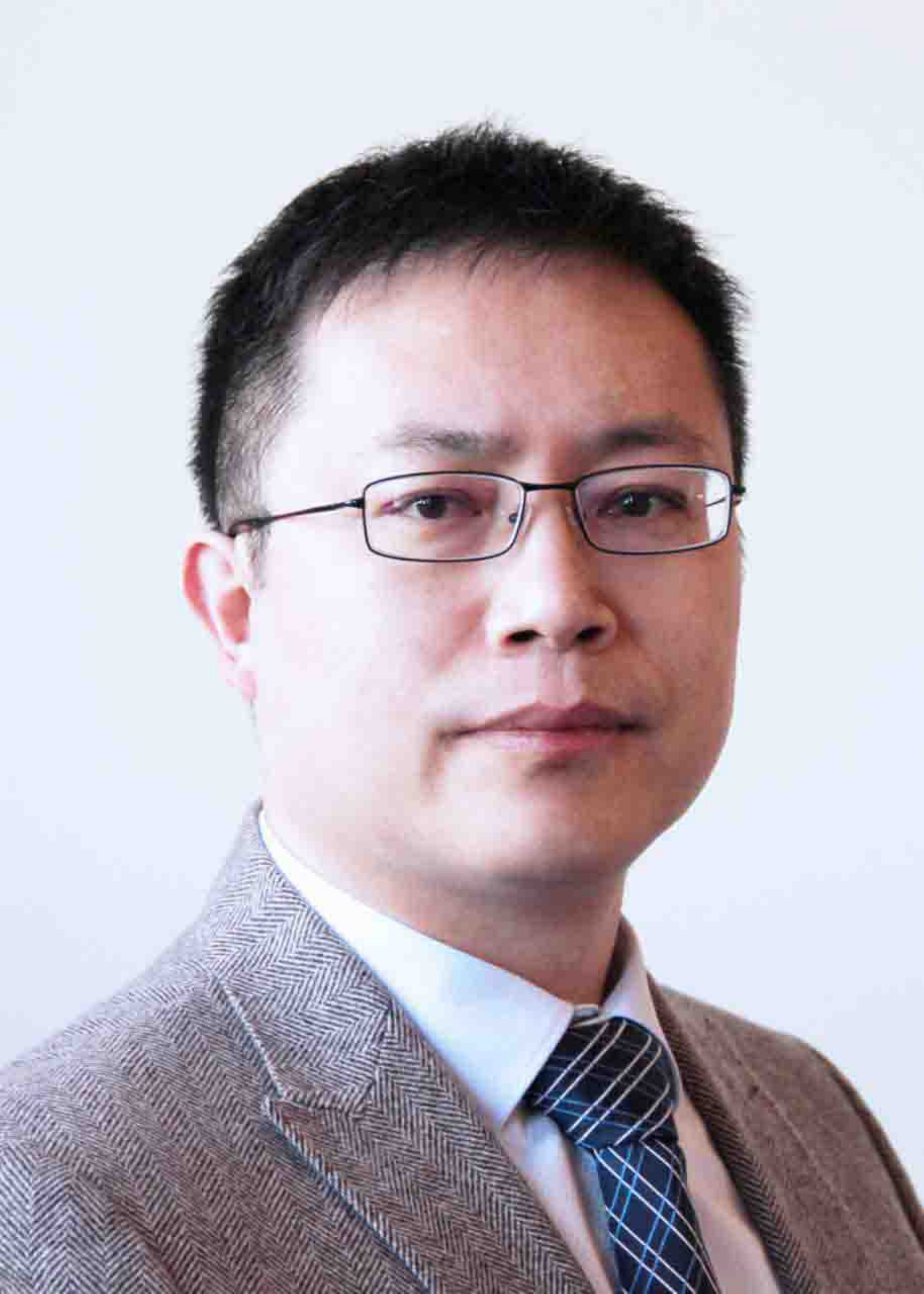}}]{Cheng-Xiang Wang [S'01-M'05-SM'08-F'17]} received the BSc and MEng degrees in Communication and Information Systems from Shandong University, China, in 1997 and 2000, respectively, and the PhD degree in Wireless Communications from Aalborg University, Denmark, in 2004.

He was a Research Assistant with the Hamburg University of Technology, Hamburg, Germany, from 2000 to 2001, a Visiting Researcher with Siemens AG Mobile Phones, Munich, Germany, in 2004, and a Research Fellow with the University of Agder, Grimstad, Norway, from 2001 to 2005. He has been with Heriot-Watt University, Edinburgh, U.K., since 2005, where he was promoted to a Professor in 2011. In 2018, he joined Southeast University, China, as a Professor. He is also a part-time professor with the Purple Mountain Laboratories, Nanjing, China. He has authored three books, one book chapter, and more than 370 papers in refereed journals and conference proceedings, including 23 Highly Cited Papers. He has also delivered 18 Invited Keynote Speeches/Talks and 7 Tutorials in international conferences. His current research interests include wireless channel measurements and modeling, B5G wireless communication networks, and applying artificial intelligence to wireless communication networks.

Prof. Wang is a fellow of the IET, an IEEE Communications Society Distinguished Lecturer in 2019 and 2020, and a Highly-Cited Researcher recognized by Clarivate Analytics, in 2017-2019. He is currently an Executive Editorial Committee member for the IEEE TRANSACTIONS ON WIRELESS COMMUNICATIONS. He has served as an Editor for nine international journals, including the IEEE TRANSACTIONS ON WIRELESS COMMUNICATIONS from 2007 to 2009, the IEEE TRANSACTIONS ON VEHICULAR TECHNOLOGY from 2011 to 2017, and the IEEE TRANSACTIONS ON COMMUNICATIONS from 2015 to 2017. He was a Guest Editor for the IEEE JOURNAL ON SELECTED AREAS IN COMMUNICATIONS, Special Issue on Vehicular Communications and Networks (Lead Guest Editor), Special Issue on Spectrum and Energy Efficient Design of Wireless Communication Networks, and Special Issue on Airborne Communication Networks. He was also a Guest Editor for the IEEE TRANSACTIONS ON BIG DATA, Special Issue on Wireless Big Data, and is a Guest Editor for the IEEE TRANSACTIONS ON COGNITIVE COMMUNICATIONS AND NETWORKING, Special Issue on Intelligent Resource Management for 5G and Beyond. He has served as a TPC Member, TPC Chair, and General Chair for over 80 international conferences. He received ten Best Paper Awards from IEEE GLOBECOM 2010, IEEE ICCT 2011, ITST 2012, IEEE VTC 2013-Spring, IWCMC 2015, IWCMC 2016, IEEE/CIC ICCC 2016, WPMC 2016, and WOCC 2019. E-mail: chxwang@seu.edu.cn.
\end{IEEEbiography}

\begin{IEEEbiography}[{\includegraphics[width=1in,height=1.25in,clip,keepaspectratio]{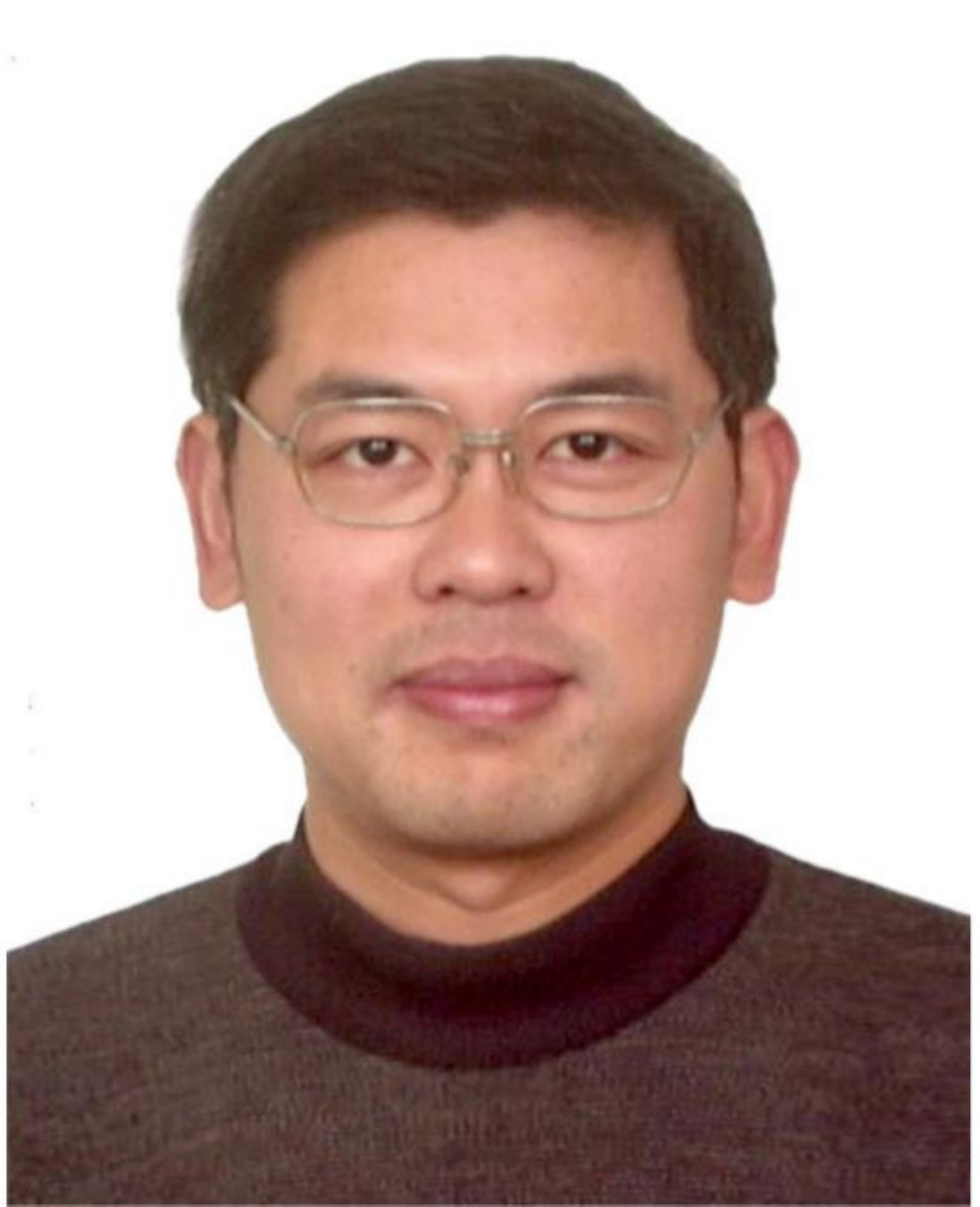}}]
{Xuming Fang} [SM'16] received the B.E. degree
in electrical engineering in 1984, the M.E. degree
in computer engineering in 1989, and the PhD
degree in communication engineering in 1999 from
Southwest Jiaotong University, Chengdu, China. He
was a Faculty Member with the Department of
Electrical Engineering, Tongji University, Shanghai,
China, in September 1984. He then joined the School
of Information Science and Technology, Southwest
Jiaotong University, Chengdu, where he has been a
Professor since 2001, and the Chair of the Department of Communication Engineering since 2006. He held visiting positions
with the Institute of Railway Technology, Technical University at Berlin,
Berlin, Germany, in 1998 and 1999, and with the Center for Advanced
Telecommunication Systems and Services, University of Texas at Dallas,
Richardson, in 2000 and 2001. He has, to his credit, around 200 highquality research papers in journals and conference publications. He has
authored or coauthored five books or textbooks. His research interests include
wireless broadband access control, radio resource management, multi-hop
relay networks, and broadband wireless access for high speed railway. Dr. Fang is the Chair of IEEE Vehicular Technology Society of Chengdu Chapter. Email: xmfang@swjtu.edu.cn.
\end{IEEEbiography}

\begin{IEEEbiography}[{\includegraphics[width=1in,height=1.25in,clip,keepaspectratio]{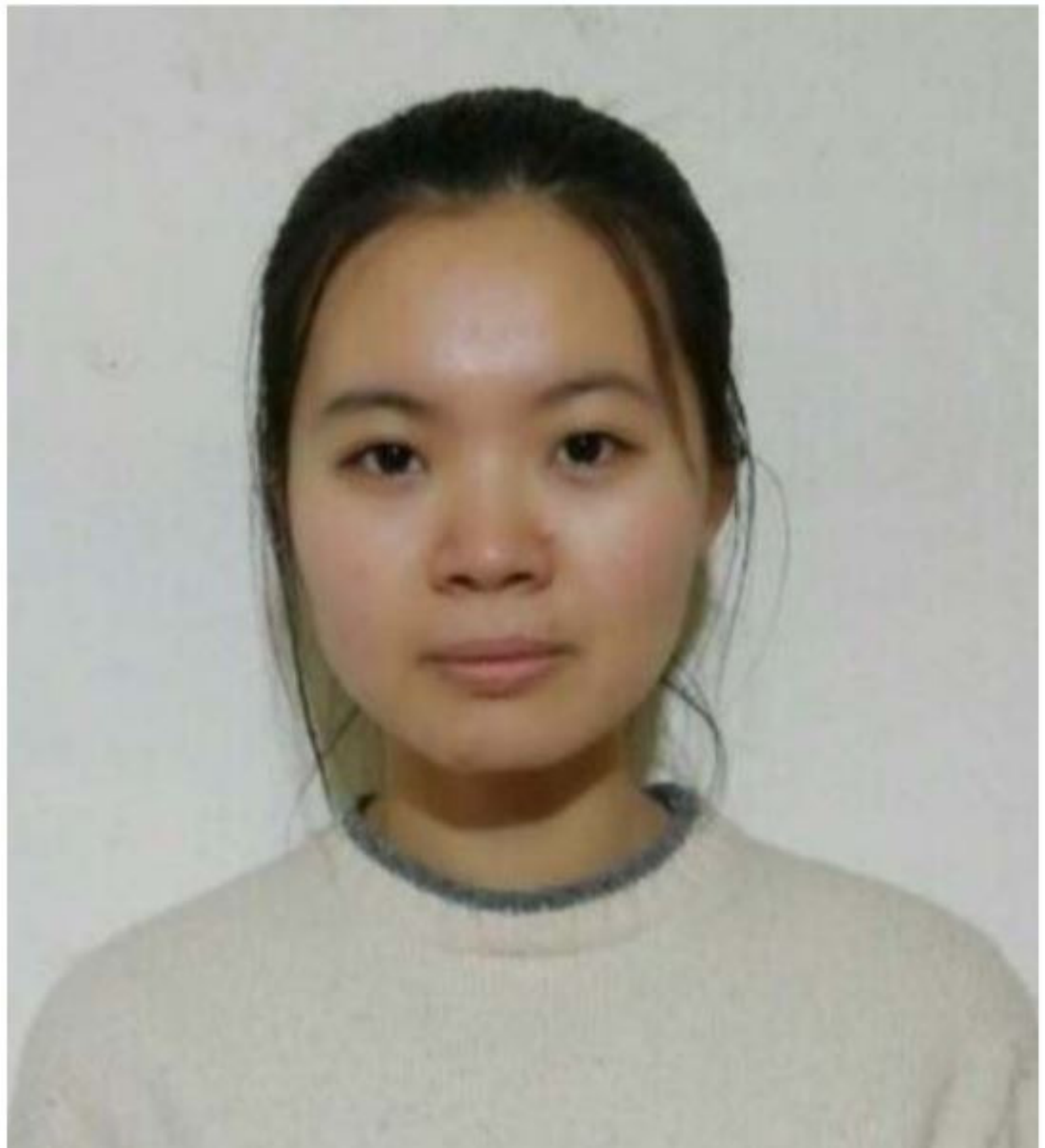}}]
{Li Yan} is a lecture at Southwest Jiaotong University, China, where she received the B.E. degree in communication engineering in 2012 and the Ph.D
degree in communication and information systems in
2018. She was a visiting student in the Department
of Electrical and Computer Engineering, University
of Florida, USA from Sept. 2017 to Sept. 2018.
Her research interests include 5G communications,
mobility managements, network architecture, millimeter wave communications, and HSR wireless
communications. Email: liyan12047001@my.swjtu.edu.cn.
\end{IEEEbiography}

\begin{IEEEbiography}[{\includegraphics[width=1in,height=1.25in,clip,keepaspectratio]{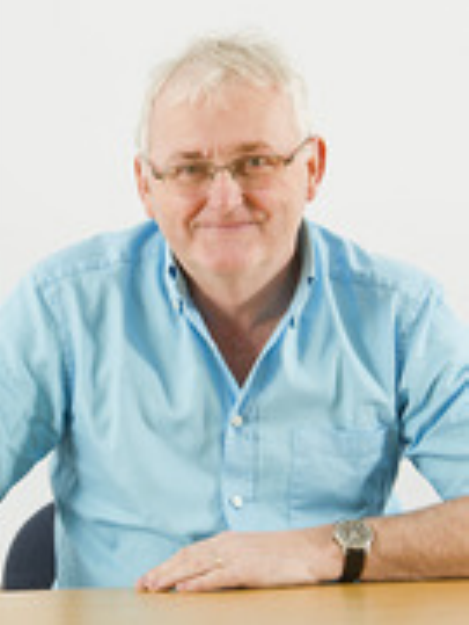}}] {Stephen McLaughlin} [F'11] received his PhD degree from the University of Edinburgh in 1989. He has been the Head of School of Engineering and Physical Sciences at Heriot-Watt University since 2011. He is a Fellow of the Royal Academy of Engineering, Royal Society of Edinburgh, and IET. Email: s.mclaughlin@hw.ac.uk.
\end{IEEEbiography}

\EOD

\end{document}